\documentclass[journal=jacsat,manuscript=article]{achemso}

\usepackage[version=3]{mhchem} 
\usepackage{subfigure}
\usepackage{amsmath}
\usepackage{amssymb}
\usepackage{graphicx}
\usepackage[final, defaultcolor=red]{changes}
\usepackage{comment}
\usepackage{xspace}
\usepackage[symbol]{footmisc}

\newcommand{\NLP}{\textit{PolymerSolarCells$_{NLP}$\xspace}}
\newcommand{\Curated}{\textit{PolymerSolarCells$_{Curated}$\xspace}}
\author{Pranav Shetty}
\affiliation{School of Computational Science \& Engineering, Georgia Institute of Technology, 771 Ferst Drive NW, Atlanta, Georgia 30332, USA}
\author{Aishat Adeboye}
\affiliation{School of Chemical \& Biomolecular Engineering, Georgia Institute of Technology, 771 Ferst Drive NW, Atlanta, Georgia 30332, USA}
\author{Sonakshi Gupta}
\affiliation{School of Materials Science and Engineering, Georgia Institute of Technology, 771 Ferst Drive NW, Atlanta, Georgia 30332, USA}
\author{Chao Zhang}
\affiliation{School of Computational Science \& Engineering, Georgia Institute of Technology, 771 Ferst Drive NW, Atlanta, Georgia 30332, USA}
\author{Rampi Ramprasad}
\affiliation{School of Materials Science and Engineering, Georgia Institute of Technology, 771 Ferst Drive NW, Atlanta, Georgia 30332, USA}
\email{rampi.ramprasad@mse.gatech.edu}

\title[Structured Information from Material Science Literature]
  {Accelerating materials discovery for polymer solar cells: Data-driven insights enabled by natural language processing}

\begin{document}

\begin{abstract}

We present a \replaced{simulation of various active learning strategies for the discovery of polymer solar cell donor/acceptor pairs using data extracted from the literature spanning $\sim$ 20 years by a natural language processing pipeline}{natural language processing pipeline that was used to extract polymer solar cell property data from the literature and simulate various active learning strategies}. While data-driven methods have been well established to discover novel materials faster than Edisonian trial-and-error approaches, their benefits have not been quantified \added{for material discovery problems that can take decades}. Our approach demonstrates a potential reduction in discovery time by approximately 75 \%, equivalent to a 15 year acceleration in material innovation. Our pipeline enables us to extract data from greater than 3300 papers which is $\sim$5 times larger \added{and therefore more diverse} than similar data sets reported by others. We also trained machine learning models to predict the power conversion efficiency and used our model to identify promising donor-acceptor combinations that are as yet unreported. We thus demonstrate a \replaced{pipeline}{workflow} that goes from published literature to extracted material property data which in turn is used to obtain data-driven insights. Our insights include active learning strategies that can \replaced{be used to}{simultaneously optimize the material system and} train strong predictive models of material properties \added{or be robust to the initial material system used}. This work provides a valuable framework for \added{data-driven} research in materials science.




\end{abstract}
\section{Introduction}

Machine learning (ML) methods have become ubiquitous in materials science. Indeed, data-driven methods have enabled the discovery of new materials for applications such as high-breakdown strength dielectric polymers \cite{mannodi2018scoping, ma2015rational, wu2021dielectric}, heussler alloys \cite{he2022computationally, jia2022unsupervised}, organic photovoltaics with high power conversion efficiency \cite{kranthiraja2022machine, sun2019machine} and gas separation membranes with high selective permeability\cite{yang2022machine}. Active learning is a technique that is used to drive this improvement. It leverages trained models of material properties to identify promising candidate materials and then augment the training set once the candidate material is ``measured", to improve the predictive performance of the ML models. 
Active learning methods have been used to discover high hole mobility thin films \cite{macleod2020self}, alloys for gas turbine engine blades\cite{khatamsaz2022multi}, and high glass transition temperature polymers \cite{kim2019active}. 
Recent work has benchmarked the performance of various active learning strategies on problems such as electrocatalysis \cite{rohr2020benchmarking}, magnetic properties\cite{wang2022benchmarking}, and band gap \cite{borg2023quantifying}. 
\replaced{Several studies have also benchmarked the performance of human scientists against bayesian optimization algorithms for the optimization of materials systems such as polyoxometallate clusters\cite{humanrobot2017, humanrobot2019}, and palladium-catalyzed arylations \cite{shields2021bayesian}. However, these studies were laboratory-scale experiments that were run under a controlled setting over a short time scale relative to the many decades it can take for materials discovery to play out for important applications.}{However, no study has compared the time taken to experimentally determine an optimal material system with desirable properties through trial and error against the time taken to discover the same material system using machine learning methods.} Polymer solar cells offer a promising playground for us to quantify the time savings offered by the use of machine learning methods in materials science using historical data.

Traditionally solar cells are inorganic and are made out of either crystalline silicon or amorphous silicon. Light incident on the surface of the solar cell produces electron-hole pairs which generates a potential that can drive an external load. Due to the use of an inorganic material like silicon, the cost of manufacturing such cells is increased due to the high temperatures required to process silicon. Polymer solar cells, on the other hand, are made of organic materials and can thus be processed at much lower temperatures. 
The most common configuration of a polymer solar cell is as a bulk hetero-junction blend of two materials known as the donor and acceptor. 
Finding optimal donor/acceptor combinations that lead to high power conversion efficiency (PCE) is an active area of research. Historically, the acceptor was based on fullerenes, typically [6,6] phenyl-C61-butyric acid methyl ester (PCBM). In recent years, though non-fullerene acceptors have gained traction and have led to solar cell configurations with much higher efficiency \cite{fu2019polymer}.

The chemical space for donors and acceptors is vast, making machine learning methods a promising avenue for narrowing down the chemical space for testing candidate materials. Past work has developed machine learning models for polymer solar cells to identify promising donors for fullerene systems \cite{nagasawa2018computer} as well as promising non-fullerene acceptors \cite{miyake2021machine}. Ref. \citenum{sun2019machine} uses trained ML models of PCE to find promising candidate materials and then tests cells fabricated using ML predictions to discover new viable candidates. The number of experimental papers published in this space is quite large however and covers a wide chemical space. In this work, we found over $\sim$ 3300 relevant papers using our pipeline. The papers that typically applied ML methods to polymer solar cells, relied on data that was manually collected and covered about 500 papers on average. A summary of publicly available polymer solar cell data sets is provided in Table \ref{tab:psc_datasets}. Thus, due to the manual nature of data collection, nearly five-sixth of relevant papers typically cannot be parsed for data extraction due to time constraints. The use of natural language processing (NLP) offers a potential solution to the problem of material property data collection at scale.

\begin{table}[!h]
\centering
\caption{Comparing public data sets of polymer solar cell device characteristics. \added{\textit{PSC} in this table refers to \textit{PolymerSolarCells}}.}
\begin{tabular}{|p{4cm}|p{5.3cm}|p{3.6cm}|c|}
\hline
Type of data & Number of \added{unique} data points & Number of papers & Reference \\
\hline
Fullerenes only & 1200 & 500 & \citenum{nagasawa2018computer} \\
\hline
Non-fullerenes only & 1001 & - & \citenum{greenstein2023screening} \\
\hline
Non-fullerenes only & 1318 & 558 & \citenum{kranthiraja2022machine, miyake2021machine} \\
\hline 
Fullerenes only & 350 & - & \citenum{lopez2016harvard} \\
\hline
Fullerenes and non-fullerenes & \textit{PSC$_{NLP}$}: 2585 \par \textit{PSC$_{Curated}$}: 867 & \textit{PSC$_{NLP}$}: 3307 \par \textit{PSC$_{Curated}$}: 861 & This work \\
\hline
\end{tabular}
\label{tab:psc_datasets}
\end{table}

NLP is a way to get computers to understand human text. NLP has been used in materials science to obtain insights from inorganic literature as well as polymer literature by extracting structured material property data or synthesis data from a large number of papers \cite{tshitoyan2019unsupervised, kononova2019text, guo2021automated, court2018auto, jensen2019machine}. We used NLP methods to extract device characteristics of polymer solar cells from the abstracts of materials science literature. A data extraction pipeline was developed as part of a previous work \cite{shettyautomated, shetty2021machine, shetty2023general}, that recognized the categories of words in text using machine learning models and used heuristic rules to link them into a material property record. Using this pipeline, we obtained a data set of donors and acceptors and their corresponding device characteristics such as PCE which we call \NLP. We curated a subset of \replaced{\NLP{}}{the NLP extracted data} to record the structure of the donor and acceptor as a SMILES string\cite{weininger1989smiles} \added{which we call \Curated{}}.  

\begin{figure}[!h]
    \centering
    \includegraphics[scale=0.135]{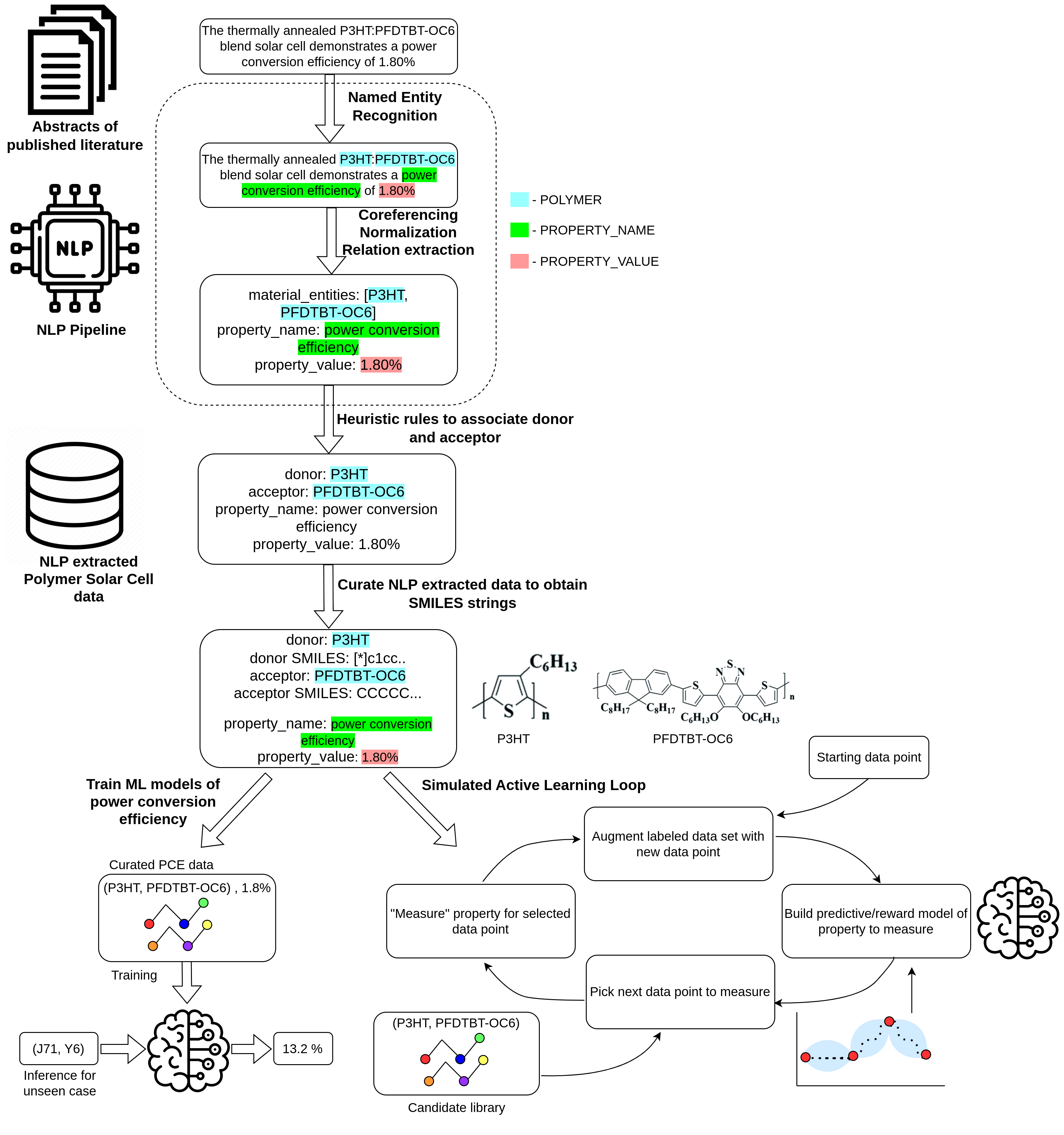}
    \caption{Pipeline used for extracting polymer solar cell PCE data from published literature which is then used in two ways 1) To train ML models of PCE and predict high-performing donor/acceptor pairs not reported in the literature and 2) To simulate an active learning loop through which donor/acceptor pairs are `discovered' sequentially. (J71, Y6) referenced in the figure is a donor/acceptor pair.}
    \label{fig:workflow}
\end{figure}

Using \replaced{\Curated}{the data curated by us}, we trained a machine learning model that takes the donor and acceptor as input and predicts the PCE (overall pipeline shown in Figure \ref{fig:workflow}). We used this trained model to predict the PCE for all donor/acceptor combinations not found in \replaced{\Curated}{our curated data set} to identify promising candidates with high PCE. This training methodology was also used while simulating an active learning loop. \replaced{}{Comparing active learning generated paths of donor/acceptor systems against the time trend of PCE in the literature enables us to estimate the time that would have been saved, had data-driven methods been used from the start to drive materials discovery.} The fundamental premise of using machine learning in materials science is that we can discover new material systems faster than would otherwise be possible \added{through trial-and-error}. We provide the first quantitative evidence for this \added{that goes beyond laboratory scale active learning experiments} by comparing the \replaced{sequence of donor/acceptor systems}{paths} obtained using active learning to the actual evolution of PCE for polymer solar cells \replaced{over a 20 year time span}{reported in the literature}. We thus make a stronger policy case for wider systemic adoption of machine learning for the discovery of new materials \added{and provide some best practices for the deployment of active learning for materials discovery}. 

\section{Methods}
\label{sec:methods}

\subsection{Data extraction pipeline from literature}

We created a corpus of 2.4 million materials science journal articles by web scraping and by downloading from publisher-specific APIs. Elsevier, Wiley, Royal Society of Chemistry, American Chemical Society, Springer Nature, Taylor \& Francis, and the American Institute of Physics are the publishers included in our corpus \cite{shettyautomated}. These papers span the years 2000 to 2022. A subset of 750 abstracts was annotated using an ontology specific to the materials science domain consisting of the entity types POLYMER, POLYMER\_CLASS, PROPERTY\_VALUE, PROPERTY\_NAME, MONOMER, ORGANIC\_MATERIAL, INORGANIC\_MATERIAL, and MATERIAL\_AMOUNT. A Named Entity Recognition (NER) model was trained using the annotated data set. The trained NER model along with heuristic rules for entity linking was used to extract material property data from the abstracts of all papers in our corpus that were polymer-relevant. This extracted data set represents all material property data reported in the abstracts of our corpus of papers. A subset of this material property data corresponds to polymer solar cells and the selection and curation of this subset is described next. Refer to Ref. \citenum{shetty2023general} for further details.

\subsection{Creating \replaced{\NLP{}}{the polymer solar cells data set}}
Polymer solar cells \footnote{\added{Note that our usage of polymer solar cells refers to cells in which at least the donor is a polymer \cite{lee2020eco, sorrentino2021interlayers, jin2023recent},  in contrast to all-polymer cells \cite{yin2020recent, ma2022polymer} in which the donor and acceptor are polymers.}} was the domain with the largest number of papers in our corpus. It is \replaced{}{also} conventional for scientists to record information such as the donor/acceptor system developed in the paper as well as its key device characteristics in the abstract of the paper. Moreover, this field has for the most part developed over the $\sim$20 year time span between 2000 and 2022 which overlaps well with the time frame of our corpus of papers. 
Most of the research in this field has also focused on developing a donor/acceptor that improves one key property namely the power conversion efficiency of the resulting cell. This makes it an ideal playground for our active learning experiments.

We picked the subset of abstracts (from which we extracted all material property data), that contained the keyword ``solar cell" and contained reported values for PSC device characteristics. 
We also excluded keywords such as ``perovskite", ``dye-sensitized", ``tandem solar cell", ``quantum dot", ``hybrid solar cell", ``silicon solar cell" and ``ternary solar cell". 
In practice, it is common for authors working in polymer solar cells to report the device characteristics associated with the most important donor-acceptor combinations tested in the paper in the abstract \added{as they are the most significant results reported in the paper}. \replaced{}{Most papers report innovations in the materials used in the cell and its corresponding performance which are typically the most important results reported in the paper and are hence mentioned in the abstract.}
Since the ontology used in Ref. \citenum{shetty2023general} was not specific to polymer solar cells, it was necessary to distinguish donor and acceptor labels for material entities reported in the abstract.
We labeled material entities as donors or acceptors based on the following hierarchy of rules:

\begin{enumerate}
    \item Donors and acceptors are often separated by a ``/" or a ``:" such as ``P3HT/PCBM" or ``P3HT:PCBM" wherein the first entry is the donor and the second entry is the acceptor. Material entities mentioned in this manner were \replaced{labeled}{marked} as donors and acceptors respectively
    \item POLYMER or ORGANIC entities that co-occur with the word donor or acceptor were \replaced{labeled}{counted} as the donor and acceptor respectively.
    \item If PCBM or any \replaced{other}{coreferent of a} fullerene is mentioned in the text, then it was labeled as the acceptor.
    \item If an ORGANIC entity is present in the list of materials for a material property record, then it was \replaced{labeled}{treated} as the acceptor and the POLYMER entity in that material list was \replaced{labeled}{treated} as the donor. \added{This case could also correspond to a small molecule donor and a polymeric acceptor. However, small molecule donors are rarely observed in our dataset and are hence ignored in this rule.}
\end{enumerate}

Anything that fell outside the above set of rules was marked as being ambiguous and required manual curation.

It is common for polymers to be referred to by generic identifiers such as P1 or P2 and for the corresponding structure to be provided in the paper. In such cases, we changed the name of the entity to P1\_DOI where DOI is the digital object identifier of the paper. This helps distinguish this case from other papers where similar identifiers were used. \added{\NLP{} thus contains NLP extracted donor/acceptor pairs and their corresponding device characteristics. The NLP pipeline also extracts coreferents (such as abbreviations) for the donor and acceptor.}

\subsection{\replaced{Curating \NLP{} to create \Curated}{Data curation}}
To \replaced{}{use the above data set to} train machine learning models, we curated \replaced{a subset of \NLP{}}{the NLP extracted data further} to 1) fix any \replaced{extraction errors}{mistakes} made by the NLP pipeline and 2) collect SMILES strings for donors and acceptors which can be used to generate a structural fingerprint of the material entity for training models. \added{The SMILES string is a string representation of the structure of a molecule.}

\added{Using each extracted datapoint in \NLP{} as the starting point, curators checked the corresponding abstract to see if the NLP extracted data was correct. If any part of the NLP extracted data was incorrect, then annotators fixed the entities extracted incorrectly. This included ensuring that the donor, acceptor, and corresponding device characteristics mentioned in the abstract i.e. PCE, fill factor, open circuit voltage, and short circuit were recorded correctly. Any relevant material property data present in the abstract that the NLP pipeline failed to extract was recorded manually by the annotators. There were cases where only the donor or the acceptor was mentioned in the abstract while the other had to be located in the body of the paper. In such cases, the material entity not found in the abstract is provided in single quotes to distinguish it from cases where the material entity is provided in the abstract. The SMILES strings for all donors and acceptors were also recorded. The SMILES strings for the most common donors and acceptors found in our corpus were first created and then pre-populated in the corresponding row to avoid repeating the effort of recreating those SMILES strings. For every other case, the SMILES string was constructed by looking for the structure among the figures in the paper and using the draw tool found in \url{polymergenome.org} to construct the SMILES string. In cases where the structure was not found in the paper, the annotators looked for supporting references that contained the structure. Note that we used the psmiles convention as defined in Ref. \citenum{kuenneth2023polybert} for polymer SMILES strings. Creating the SMILES strings was the most time-consuming aspect of data curation.} 

\added{In addition to the basic curation workflow described above, we employed several filtering criteria.} We were only interested in properties reported for bulk heterojunction polymer solar cells. Thus the donor in each data point is a polymer and papers dealing with ternary systems, i.e., with multiple donors or multiple acceptors were excluded. \added{This was done so that the learning problem could be formulated using a single donor and acceptor. Small molecule donors were excluded as very few papers in \NLP{} report such data given that our NLP pipeline was engineered to extract polymer property data. Out of 150 papers randomly sampled from \NLP{}, there was only one paper corresponding to a small molecule donor.} We also excluded papers reporting properties computed through simulations or materials that were discovered using data-driven methods such as Ref. \citenum{sun2019machine}. 

The device characteristics reported in \replaced{\Curated{}}{the data set} are power conversion efficiency, fill factor, open circuit voltage, and short circuit current.
In the analysis that follows, we only make use of the power conversion efficiency\replaced{}{, the complete data set is released as part of this work}. Note that all property values curated were mentioned in the abstract. \added{In case multiple property values are found in the abstract corresponding to different fabrication conditions, the annotators were instructed to record only those corresponding to the highest power conversion efficiency. This corresponds to the optimal fabrication condition.}
\replaced{}{There were cases where only the donor or acceptor was mentioned in the abstract while the other had to be located in the body of the paper. In such cases, the material entity not found in the abstract is provided in single quotes to distinguish it from cases where the material entity is provided in the abstract.}

\subsection{Machine learning prediction of power conversion efficiency}

We used \replaced{\Curated{}}{the curated data set} to build machine learning models of the power conversion efficiency (PCE) of a polymer solar cell. We modeled the PCE as being dependent on only the donor and acceptor. In practice, the PCE depends on factors like the weight fractions of the donor and acceptor and other device fabrication parameters \cite{zhao2020recent}. We however assumed that the property values reported in the abstract are for systems in which other factors have been optimized. \added{This is because scientists typically report the highest PCE value they measure as one of the key contributions of their paper which in turn corresponds to an optimized cell. We empirically validated this assumption by sampling 20 papers in \Curated{} and verified that this assumption held true in all 20 papers.} Our problem formulation is more general than several others reported in the literature which only use the structure of the donor and assume that the acceptor is a fullerene \cite{nagasawa2018computer} or do not consider the effect of the acceptor \cite{sun2019machine}. Due to the recent rise of non-fullerenes \cite{fu2019polymer}, it is necessary to include acceptors in the problem formulation \added{as done in Ref. \citenum{padula2019concurrent}.} For donor-acceptor pairs for which multiple power conversion efficiencies are available, we took the median of all available values. The values may differ due to differences in the electron or hole transport layer, additives used during the synthesis of the active layer, or the morphology of the active layer. Taking the median ``averages" out these factors.

We used hand-crafted fingerprints described in earlier work \cite{pg} to fingerprint the donor and acceptor. While other fingerprints have been developed for featurizing polymers using ideas such as graph neural networks \cite{gurnani2023polymer} and BERT-based fingerprinting \cite{kuenneth2023polybert}, we used handcrafted fingerprints for this work as they can be used to fingerprint polymers as well as organic molecules. Moreover, each fingerprint component corresponds to a pre-defined chemical feature which enables us to ``peer" into the model and determine why the model made certain predictions. In cases where the donor or acceptor is a copolymer, we take the average fingerprint of all \replaced{components}{constituent monomers} of the copolymer, in keeping with past work \cite{kuenneth2021copolymer}. \added{Further details on fingerprinting methodology can be found in the Supporting Information Section 2.} Min-max scaling is performed on all fingerprint components to ensure that they are each between 0 and 1. The vector fingerprints for the donor and acceptor were then concatenated and input to a Gaussian Process Regression (GPR) model to predict the power conversion efficiency. The fingerprint dimension is 745 while the number of data points is 835. \footnote{\added{This is lower than the total number of unique donor/acceptor systems (867) contained in \Curated{} as 32 donor/acceptor systems in \Curated have devices characteristics other than PCE such as fill factor or open circuit voltage reported in the corresponding abstracts.}} Neural networks tend to overfit when the number of data points is close to the dimension of the feature space while GPR models generalize much better in this regime making it a natural choice for our problem \cite{kamath2018neural}. We used the python package scikit-learn\cite{scikit-learn} to train GPR models. We used a Matern kernel\cite{williams2006gaussian} \added{with $\nu=1.5$. We chose this kernel as it is well tested in a variety of materials informatics problems \cite{noack2020autonomous, lookman2019active, tran2020methods}.}.
\replaced{\Curated}{The curated data} is split into 85 \% for training and 15 \% for testing. We used the root mean squared error (RMSE) and the coefficient of determination (r) to assess model performance.

\subsection{Data selection methods for simulating active learning of polymer solar cells}

\replaced{Using}{We used} \Curated{} \added{as the set of candidate materials} we picked a sequence of donor/acceptor pairs \added{using several data selection methods}. This process was continued until the data point with the highest PCE in our data set was discovered. This in effect, simulates an active learning generated path of how the field of polymer solar cells may have developed and can be used to compare against how the field developed. Let $y^{pred}_{i}$ be the predicted value of the GPR model for the $i^{\text{th}}$ test point and $\sigma_{i}$ be the corresponding uncertainty of the GPR model evaluated at that point where $i \in \{1,...,N\}$ where $N$ is the number of test points at a given step in the active learning loop. Suppose the current step in the active learning cycle is $T$. Let $Y_i$ represent the random variable corresponding to the measurement of the $i^{\text{th}}$ test point. If GPR is used to model $Y_i$, then it is known that $Y_i$ is Gaussian distributed with mean $y^{pred}_{i}$ and standard deviation $\sigma_{i}$. \replaced{We employed several different methods to pick the next data point to test, described below.}{The first four methods described below require a model to be trained at each step of the active learning loop while linear bandits do not train a predictive model.}

\begin{enumerate}
    \item \textbf{Gaussian Process-Upper Confidence Bound (GP-UCB)}: This uses the predicted value along with its uncertainty to pick points, i.e., $y^{pred}_{i}+\beta\sigma_{i}$. The parameter $\beta$ controls the trade-off of exploration versus exploitation. A higher value of $\beta$ indicates a belief that higher value points are likely to be found in regions of uncertainty while a lower value of beta indicates a belief that the model predictions should be exploited as is. We used a value of $\beta=1$ in this work \added{as it represents a balanced trade-off of exploration and exploitation \cite{srinivas2009gaussian}} 
    \item \textbf{Gaussian Process-Probability Improvement (GP-PI)}: Intuitively, this strategy looks for points that have the highest probability of being greater than some threshold $\theta$. $\mathbb{P}(Y_{i}>\theta) = 1-\Phi(\frac{\theta-y^{pred}_{i}}{\sigma_{i}}) = \Phi(\frac{y^{pred}_{i}-\theta}{\sigma_{i}})$. Here $\Phi(.)$ is the cumulative probability distribution function for a standard Gaussian. In practice, the threshold is selected to be the value of the highest $y$ sampled until that point, multiplied by some improvement fraction $1+\nu$, i.e., $\theta = \text{max}_{t \in \{1,...,T\}}\; y_t (1+\nu)$. We pick a value of $\nu=0.01$. \added{This provides a practical compromise between being too aggressive or too conservative in seeking improvements across all active learning cycles. This is enough to avoid negligible improvements between cycles but is not so large as to miss potential subtle improvements in areas close to the current best observations thus leading to consistent and gradual improvement \cite{jones1998efficient}.} After computing the probability of improvement for all test points, we pick the point with the greatest value.
    \item \textbf{Gaussian Process-Expected Improvement (GP-EI)}: In contrast to the previous strategy where we compute the probability of improvement over some threshold, in this strategy we look at the expected value of the difference between the value and the threshold. Thus we compute the quantity $\mathbb{E}[Y_{i}-\theta] = \int^{\infty}_{-\infty} (y_i - \theta)p(y_i)dy_i$. This evaluates to $E[Y_{i}-\theta] = (y^{pred}_{i} - \theta)\Phi(\frac{y^{pred}_{i}}{\sigma_{i}})+\sigma_{i}\phi(\frac{y^{pred}_{i}}{\sigma_{i}})$, where $\phi(.)$ is the probability distribution function of the standard gaussian. The threshold is computed the same way as GP-PI. After computing this quantity for all test points, we pick the point with the largest expected improvement.
    \item \textbf{Greedy acquisition}: This strategy simply looks at the highest predicted values $y^{pred}_{i}$. This is equivalent to using $\beta=0$ in GP-UCB.
    \item \textbf{Gaussian Process-Thompson Sampling (GP-TS)}: For each material system, we sample the value of the power conversion efficiency from a Gaussian distribution with mean $y^{pred}_{i}$ and standard deviation $\sigma_{i}$. We pick the material system with the highest sampled property value.
    
    \item \textbf{Linear contextual bandits}: In a multi-arm bandit setting, an agent interacts with an environment over several rounds. In contrast to the methods described above, we do not train an ML model at each round. During each round, the agent selects an action from a set of available actions (also known as ``arms" in the multi-arm bandit setting) where each action is represented by a d-dimensional feature vector (the ``context''). In this case, the set of actions is the available set of donor/acceptor systems. After taking an action, the agent receives a reward, which in this case is the value of the PCE ``measured" for a donor/acceptor system. The goal is to learn a policy (a mapping from context to action) that maximizes the expected cumulative reward over time. In linear contextual bandits, the reward function is assumed to be a linear function of the action. The agent's goal is to estimate the parameters of the linear model based on observed data and use it to make action selections that maximize expected rewards. We implemented Thompson Sampling for linear contextual bandits\cite{agrawal2013thompson}. 
    While training GPR models, the focus is on optimizing prediction accuracy, while in linear contextual bandits, the focus is on maximizing cumulative rewards. Instead of having a fixed data set used for training, contextual bandits involve online learning where the agent interacts with the environment and adapts its policy over time. Further mathematical details are provided in Supporting Information Section 1.
    \item \added{\textbf{Random}: A donor/acceptor pair is sampled uniformly from the candidate set at each step of the active learning cycle.}
\end{enumerate}

In each of the above methods, we could pick a batch $b$ of any size but in practice, we use $b=1$ since the cost of fabricating a cell with a given material system and measuring the PCE is much larger than the computational cost of the active learning loop. 

\section{Results and discussion}

\subsection{Analysis of polymer solar cell data}

There are 3307 documents in \replaced{\NLP}{the NLP extracted data set} out of which 861 ($\sim$ 26 \%) have been curated \added{to create \Curated} (detailed breakdown in Table \ref{tab:psc_dataset_breakdown}). 
The number of unique data points in \replaced{\NLP}{the full data set} is estimated based on the normalized names of the donors and acceptors and is an approximation. The donors are always polymers and hence can be normalized using the normalization workflow described in earlier work \cite{shetty2021machine}. The acceptors on the other hand can be polymers or organic molecules. For polymer acceptors, we used our polymer normalization workflow. For fullerene acceptors, given that only about four such acceptors are commonly used (PC61BM, PC71BM, C60, ICBA), we manually constructed a list of named entity variants for each of these. For other non-fullerene acceptors, we constructed an on-the-fly normalization data set using the coreferents for all non-fullerene acceptors mentioned among the curated papers. The number of total data points is greater than the number of unique data points as there are several material systems \replaced{}{for which a value has been} reported in multiple papers, usually differing in cell fabrication conditions. 
For \replaced{\Curated}{the curated data set}, the number of data points was computed using the canonical SMILES string of the donor and acceptor. The SMILES strings were canonicalized using the psmiles package described in Ref. \citenum{kuenneth2023polybert}. \added{Canonicalization is important as the same chemical structure can be represented as a valid SMILES string in multiple ways and canonicalization is used to create a unique SMILES string for a given chemical structure.}

\begin{figure}
    \centering
    \includegraphics[scale=0.435]{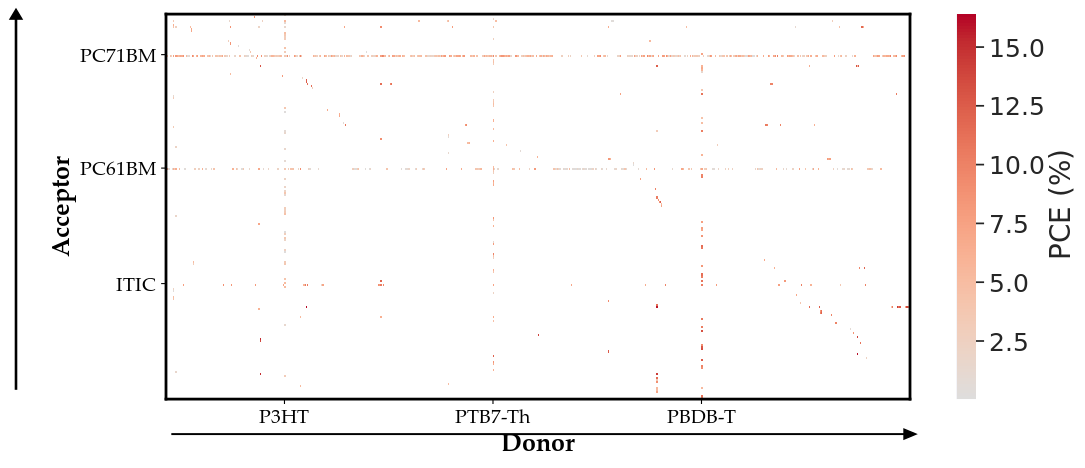}
    \caption{Entire donor/acceptor space at a glance. The donors and acceptors that are most frequently reported are shown along the axes. \added{The top three most commonly reported donors and acceptors are spaced uniformly along each corresponding axes so that they can be clearly distinguished. The remaining donors and acceptors are randomly ordered.}}
    \label{fig:donor_acceptor_space}
\end{figure}

We can see from the heatmap in Figure \ref{fig:donor_acceptor_space} that only a small fraction (0.70 \%) of the donor/acceptor space has been explored experimentally. 
\replaced{Not all}{The} donors and acceptors are \replaced{}{not} labeled as they are far too numerous. The donors on the heat map which are reported against the most acceptors correspond to P3HT (40), PTB7-Th (40), PBDB-T (37). 
The material system with the highest reported PCE for a bulk heterojunction polymer solar cell in \replaced{\Curated}{our data set} is PBDS-T as donor and BTP-ec9 as acceptor with a reported PCE of 16.4 \% \cite{bin2022efficient}.
We built a prediction pipeline that can fill in the blanks in this figure and used that to discern patterns as well as find optimal donor/acceptor combinations that would otherwise require expensive experimental measurements. A machine learning model that takes both the donor and the acceptor structure as input should in principle be able to learn correlations between the two which can be used to predict novel combinations of donors and acceptors. 


\begin{table}[!h]
\centering
\caption{Detailed composition of the NLP extracted polymer solar cell data and the curated subset}
\begin{tabular}{|p{6cm}|@{}c|c|}
\hline
Type of data & \NLP & \Curated \\
\hline
Number of papers & 3307 & 861 \\
\hline
Unique donor-fullerene pairs & 1160 & 607 \\
\hline
Total donor-fullerene data points & 2107 & 903 \\
\hline 
Unique donor- non-fullerene pairs & 1425 & 261 \\
\hline
Total donor-non-fullerene data points & 1934 & 284 \\
\hline
Number of unique donors & 1910 & 628 \\
\hline
Number of unique acceptors & 649 & 190 \\
\hline
Maximum reported power conversion efficiency & - & 16.4 \% \\
\hline
\end{tabular}
\label{tab:psc_dataset_breakdown}
\end{table}

In Table \ref{tab:nlp_ground_truth_metrics} we compare the ground truth versus the extracted data to measure the fidelity with which the NLP system can extract the relevant polymer solar cell data. 

We measured the fidelity of extraction using k-tuple metrics where $2 \leq k \leq 4$. For a 4-tuple comparison, we looked at the tuple (donor, acceptor, property\_name, property\_value). We compared the ground truth tuples against the NLP extracted tuples for each abstract to check if there is any NLP extracted tuple for which all 4 entries are identical and that is marked as a true positive (TP). If no such NLP extracted tuple is found then we have a false negative (FN). We similarly compared the NLP extracted tuples against the ground truth tuples for each abstract to check if a match is found and if not then a false positive (FP) is recorded. As the 4-tuple metric is strict, we computed 3-tuple and 2-tuple metrics as well wherein the 4-tuple is split into 3 and 6 tuples respectively for each 4-tuple, in each abstract. This allows us to see if at least some subsets of the 4-tuple are extracted correctly. This is consistent with how extraction fidelity has been measured in the literature in cases involving multiple entities in a tuple \cite{jain2020scirex}. Precision, Recall, and F1 score are then calculated as below:

\begin{equation}
\begin{split}
    \text{Precision} &= \frac{\text{TP}}{\text{TP}+\text{FP}} \\
    \text{Recall} &= \frac{\text{TP}}{\text{TP}+\text{FN}} \\
    \text{F1} &= \frac{2 \times \text{Precision} \times \text{Recall}}{\text{Precision} + \text{Recall}}
\end{split}
\end{equation}

Each of the above metrics is reported as a \% value.


\begin{table}[!h]
\centering
\caption{Comparing fidelity of NLP extracted data to ground truth data. \added{Each value in the table is a \%.}}
\begin{tabular}{|p{3cm}|@{}c|c|c|}
\hline
Metric & Precision & Recall & F1 \\
\hline
4-tuple & 46.03 & 40.25 & 42.95 \\
\hline
3-tuple & 61.34 & 55.09 & 58.05 \\
\hline
2-tuple & 72.87 & 66.18 & 69.37 \\
\hline
\end{tabular}
\label{tab:nlp_ground_truth_metrics}
\end{table}

\subsection{Predicting power conversion efficiency}

\begin{figure}
    \raggedleft
	\subfigure[]{
		\begin{minipage}{0.45\textwidth}
			\includegraphics[width=1\textwidth]{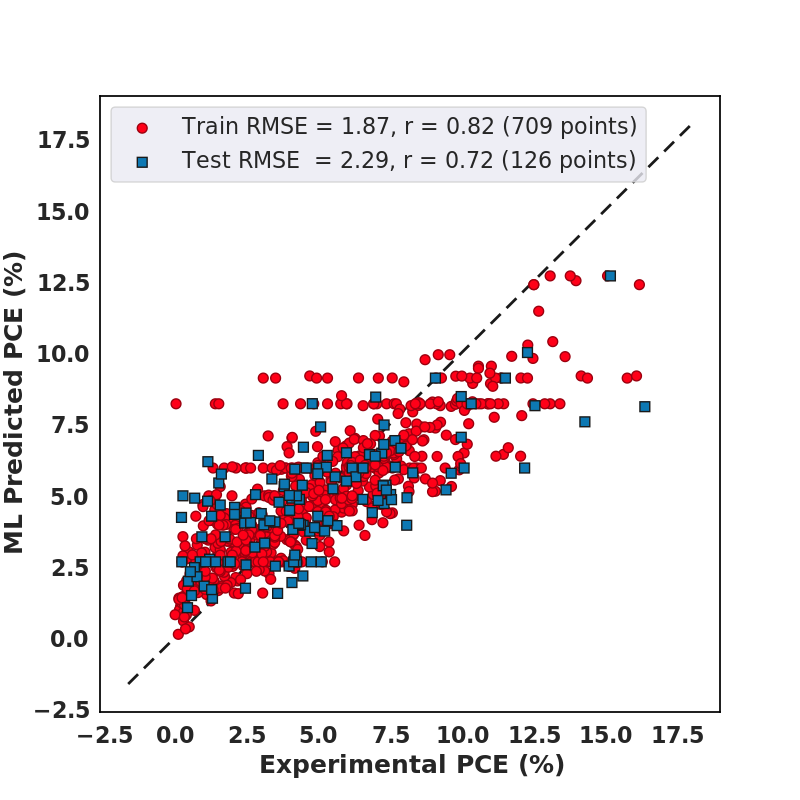}
		\end{minipage}
	}\hspace{2mm}
	\centering
        \subfigure[]{
		\begin{minipage}{0.45\textwidth}
			\includegraphics[width=1\textwidth]{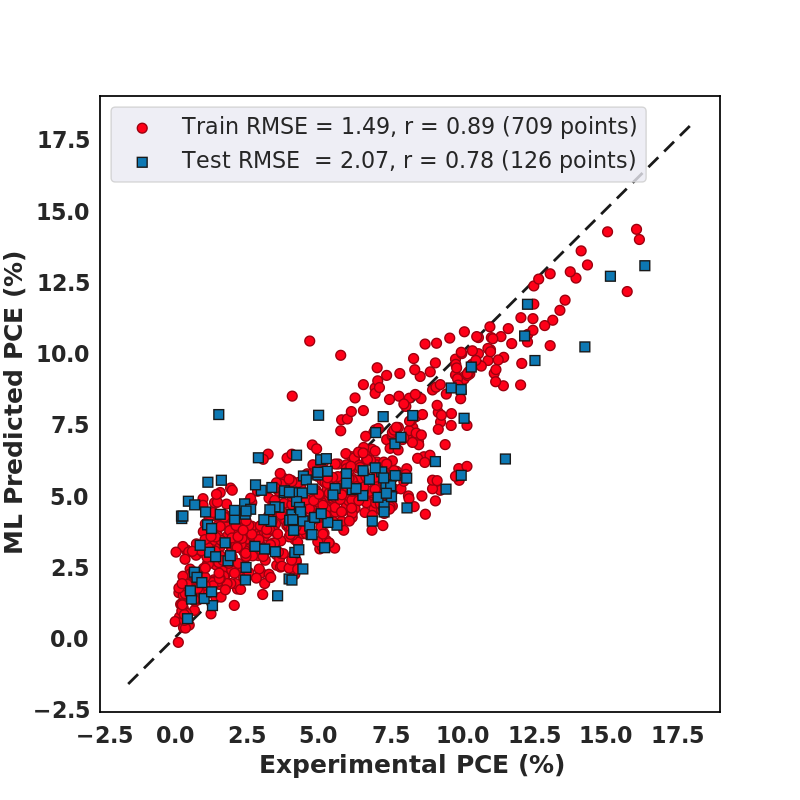}
		\end{minipage}
	}
    \caption{Parity plot for a machine learning model trained to predict power conversion efficiency. a) Model trained using only donors as input b) Model trained using donors and acceptors as input to the model}
    \label{fig:psc_parity_plot}
\end{figure}

\begin{figure}
    \centering
    \includegraphics[scale=0.435]{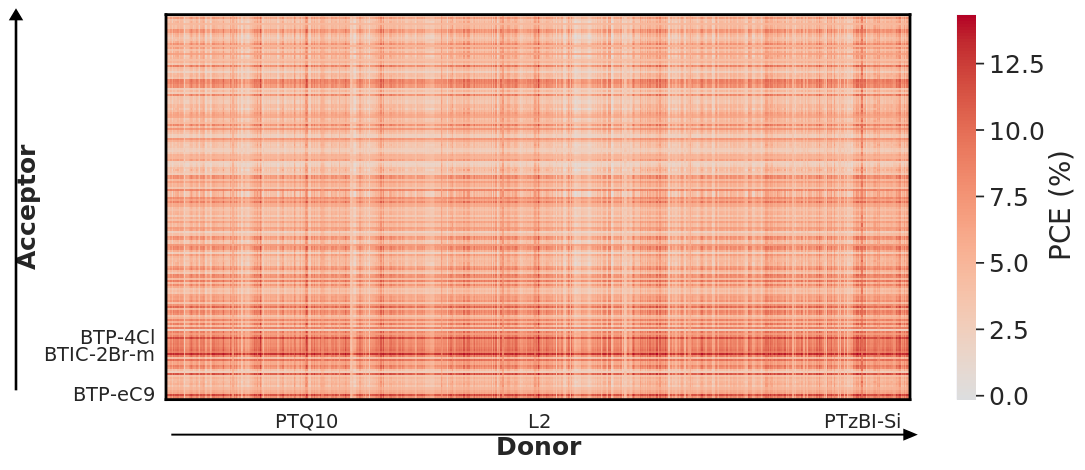}
    \caption{Predicted power conversion efficiency value for the entire donor/acceptor space. The donors and acceptors with the highest average PCE are shown along the axes. \added{The ordering of donors and acceptors is the same as Figure \ref{fig:donor_acceptor_space}.}}
    \label{fig:psc_extrapolation}
\end{figure}

The RMSE of property prediction is 2.07 \% for the model using donors as well as acceptors as seen in the parity plot in Figure \ref{fig:psc_parity_plot}. Figure \ref{fig:psc_parity_plot} compares a machine learning model trained using the donor alone and using the donor as well as the acceptor \added{as input to the model}. The donor-only data set was constructed by removing the acceptors from each data point in the donor/acceptor data set thus ensuring a fair comparison. Thus, some data points have the same donor but different values of power conversion efficiency which is due to the \added{corresponding} acceptors being different. Having additional information on the acceptor improves the prediction performance of the model by 0.22 \% as measured by the test RMSE. Note that the train-test split is identical for both Figure \ref{fig:psc_parity_plot}a and Figure \ref{fig:psc_parity_plot}b.
Figure \ref{fig:psc_extrapolation} shows the predicted power conversion efficiencies for all donor/acceptor pairs in \replaced{\Curated}{our data set}. We observe from Figure \ref{fig:psc_extrapolation} that the use of certain acceptors can improve the power conversion efficiency significantly. The bright red horizontal line at the bottom corresponds to the acceptor BTIC-2Br-m which is a Y6-based acceptor. Similarly, some of the other high-performance acceptors in Figure \ref{fig:psc_extrapolation} are also based on Y6 \cite{song2020efficiency} such as BTP-4Cl and BTP-ec9. The observations we make from this figure match well with the strategy recently employed in the polymer solar cell community of using acceptors based on Y6 and testing it with various donors. 

The highest donor/acceptor pairs predicted from our model are listed in Table \ref{tab:psc_predicted_combinations} along with the corresponding power conversion efficiency. All the acceptors are based on Y6 and the corresponding donors are thus recommendations for promising combinations. Due to the sparsity of the underlying data set, however, this extrapolation must be viewed with caution. It is possible that the acceptors with the highest recorded PCE have an undue influence on the predictions and that donor/acceptor interactions are unable to be fully captured due to the sparsity of the underlying data. The vast majority of donors are reported with just a single acceptor and likewise, most acceptors are tested with just a single donor, thus making it difficult for a model to learn the correlations effectively. To examine this possibility more closely, we looked at the top donors and acceptors from Figure \ref{fig:psc_extrapolation} by computing the average PCE over donors for each acceptor and likewise for donors (Table \ref{tab:psc_top_donor_acceptor}). \replaced{}{The top five results are shown in Table \ref{tab:psc_top_donor_acceptor}}. The top pairs reported in Table \ref{tab:psc_predicted_combinations} are not simply formed by matching the top donors with the top acceptors in Table \ref{tab:psc_top_donor_acceptor} but are non-trivial combinations of donors and acceptors, suggesting that the model may well be learning meaningful donor/acceptor co-relations. The acceptors used in the top reported pairs match closely with the top acceptors overall and the average PCE for the top acceptors is higher than that for the top donors suggesting that acceptors do indeed play a more important role than donors in determining the power conversion efficiency in the high PCE regime.

Thus, we have demonstrated the end-to-end use of NLP to facilitate the creation of a data set which is then used for building machine learning models and predicting new donor/acceptor combinations for polymer solar cells. This also validates our modeling approach for simulated active learning discussed in the next section.

\begin{table}[!h]
\centering
\caption{Donor/acceptor pairs with highest predicted power conversion efficiency}
\begin{tabular}{|c|c|p{3cm}|p{3cm}|}
\hline
Donor & Acceptor & Predicted power conversion efficiency & True PCE (if reported in literature)\\
\hline
PBDB-TF & BTIC-2Br-m & 14.34 \% & 16.1 \% \cite{wang2020bromination} \\
\hline
PM6 & BTIC-2Br-m & 14.33 \% & - \\
\hline
PFBDB-T \cite{he2019fused} & BTIC-2Br-m & 14.33 \% & - \\
\hline 
PM7 & BTP-ec9 & 14.31 \% & - \\
\hline
PM7 & BTP-4Cl & 14.30 \% & - \\
\hline
PM7 & TPIC-4Cl & 14.25 \% & 15.1 \% \cite{li2020higher} \\
\hline
PDBT-F \cite{huang2019novel} & BTIC-2Br-m & 14.20 \% & - \\
\hline
L2 & BTP-ec9 & 14.17 \% & - \\
\hline
L2 & BTP-4Cl & 14.16 \% & - \\
\hline
PBDT(T)[2F]T \cite{firdaus2017polymer} & BTIC-2Br-m & 14.07 \% & - \\
\hline
\end{tabular}
\label{tab:psc_predicted_combinations}
\end{table}

\begin{table}[!h]
\centering
\caption{Top donors and acceptors from Figure \ref{fig:psc_extrapolation}}
\begin{tabular}{|c|c|p{3cm}|p{3cm}|}
\hline
Top Donors & Average PCE (\%) & Top acceptors & Average PCE (\%)\\
\hline
PTzBI-Si \cite{zhu2019aggregation} & 8.87 & BTIC-2Br-m \cite{wang2020bromination} & 12.42 \\
\hline
L2 \cite{li2020non} & 8.68 & BTP-4Cl \cite{ma2020high} & 11.19 \\
\hline
PTQ10 \cite{sun2019achieving} & 8.46 & BTP-ec9 & 11.07 \\
\hline
PM7 \cite{zhang2020electron} & 8.44 & TPIC-4Cl \cite{li2020higher} & 10.42 \\
\hline
F13 \cite{he2020highly} & 7.98 & Y6 & 9.37 \\
\hline

\end{tabular}
\label{tab:psc_top_donor_acceptor}
\end{table}


\subsection{Simulating the `discovery' of new donor/acceptor combinations}
\begin{figure}[!h]
	\raggedleft
	\subfigure[]{
		\begin{minipage}{0.30\textwidth}
			\includegraphics[width=1\textwidth]{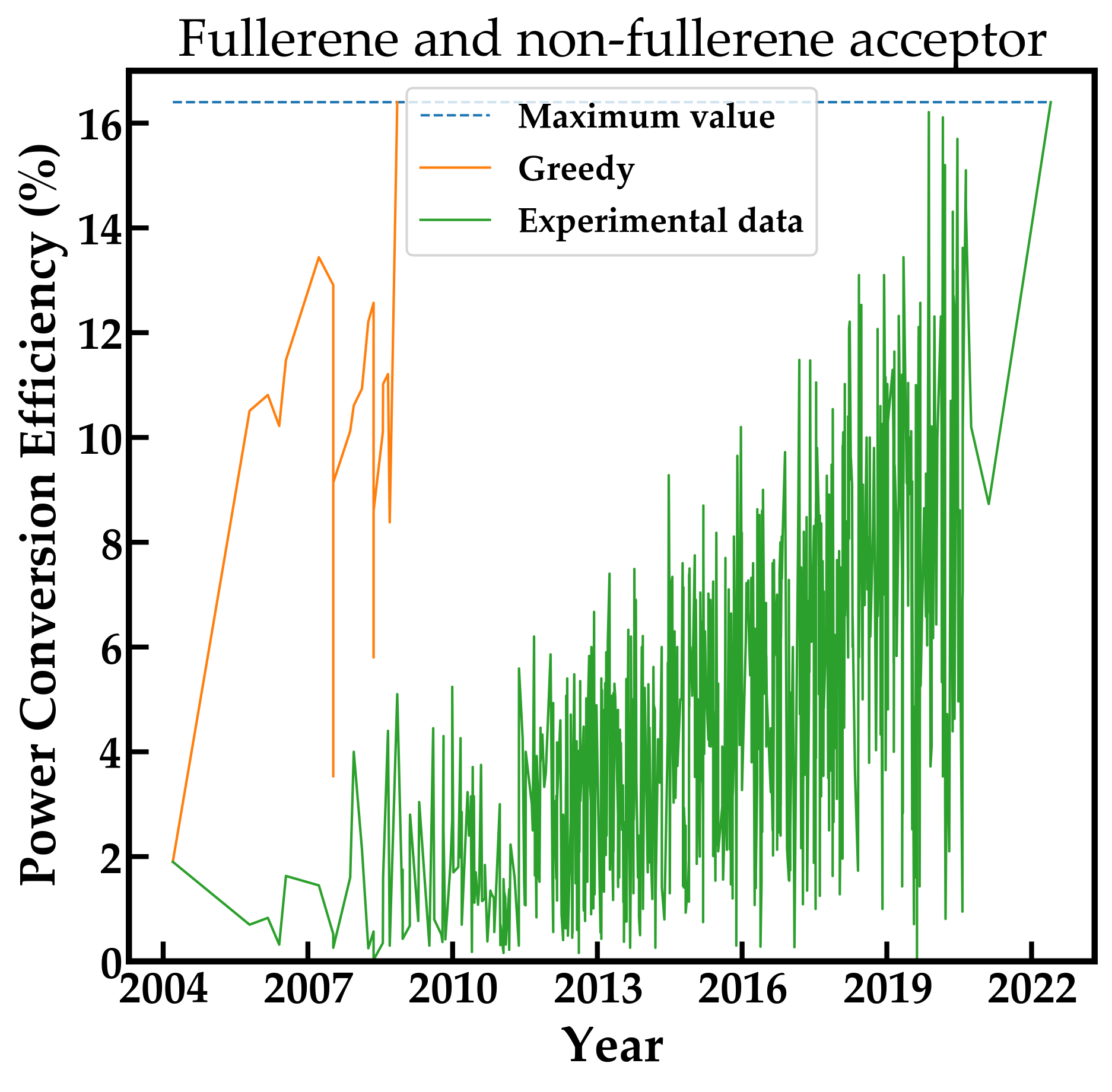}
		\end{minipage}
		\label{fig:method_comparison_both}
	}\hspace{2mm}
	\centering
        \subfigure[]{
		\begin{minipage}{0.30\textwidth}
			\includegraphics[width=1\textwidth]{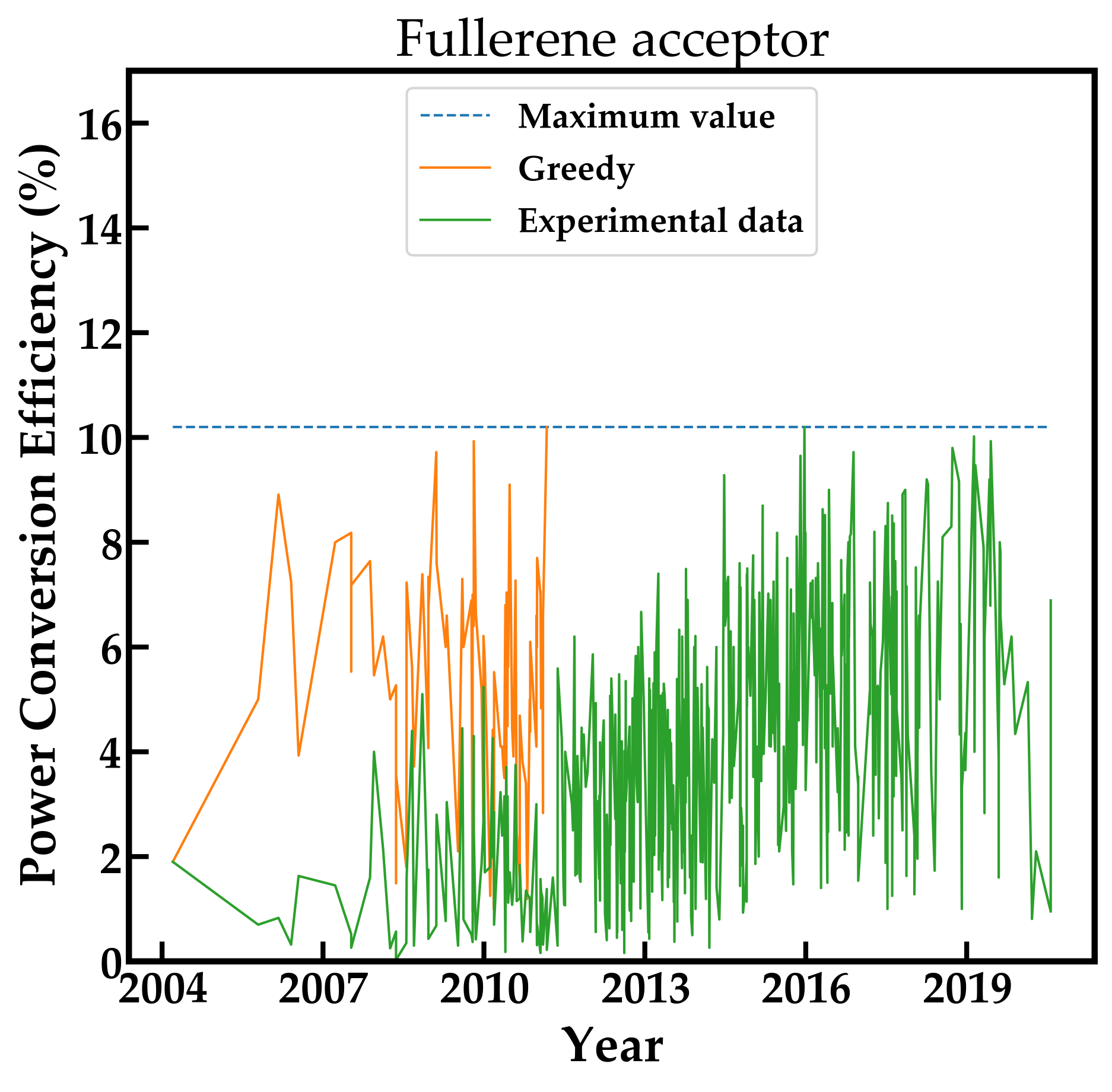}
		\end{minipage}
		\label{fig:method_comparison_FA}
	}
	\raggedright
	\subfigure[]{
		\begin{minipage}{0.30\textwidth}
			\includegraphics[width=1\textwidth]{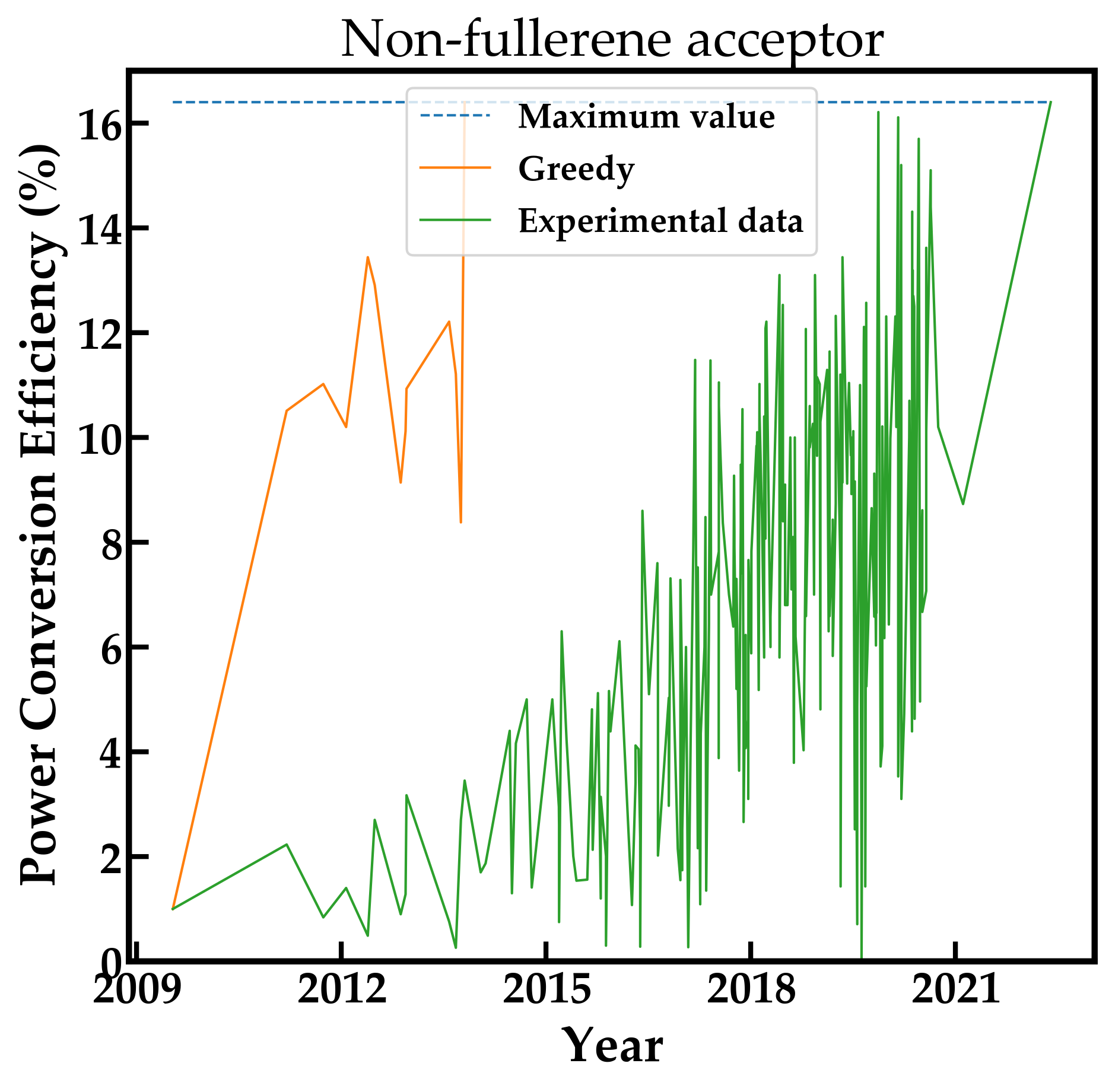}
		\end{minipage}
		\label{fig:method_comparison_NFA}
	}\hspace{2mm}
        \raggedleft
	\subfigure[]{
		\begin{minipage}{0.30\textwidth}
			\includegraphics[width=1\textwidth]{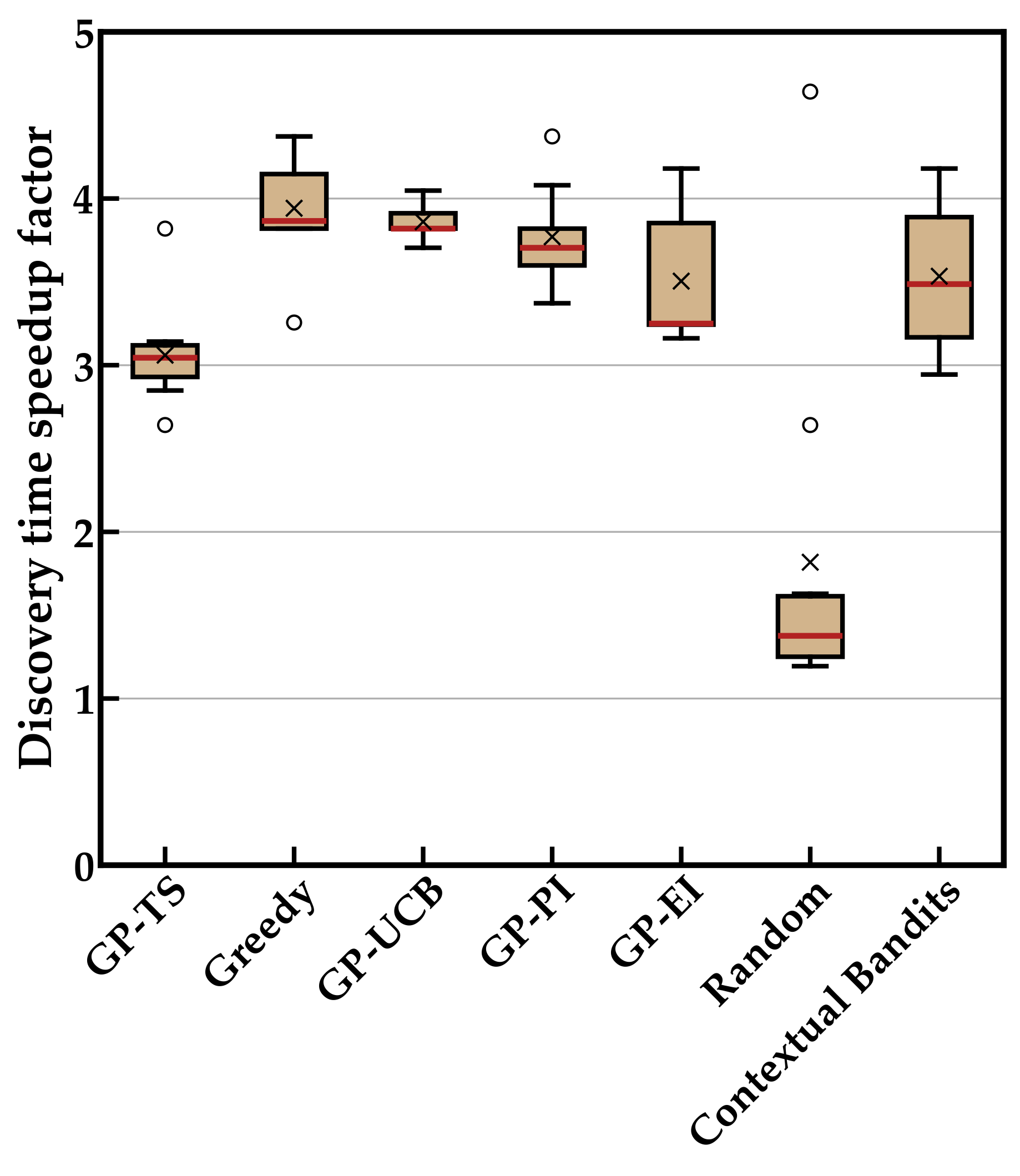}
		\end{minipage}
		\label{fig:speedup_both}
	}\hspace{2mm}
	\centering
	\subfigure[]{
		\begin{minipage}{0.30\textwidth}
			\includegraphics[width=1\textwidth]{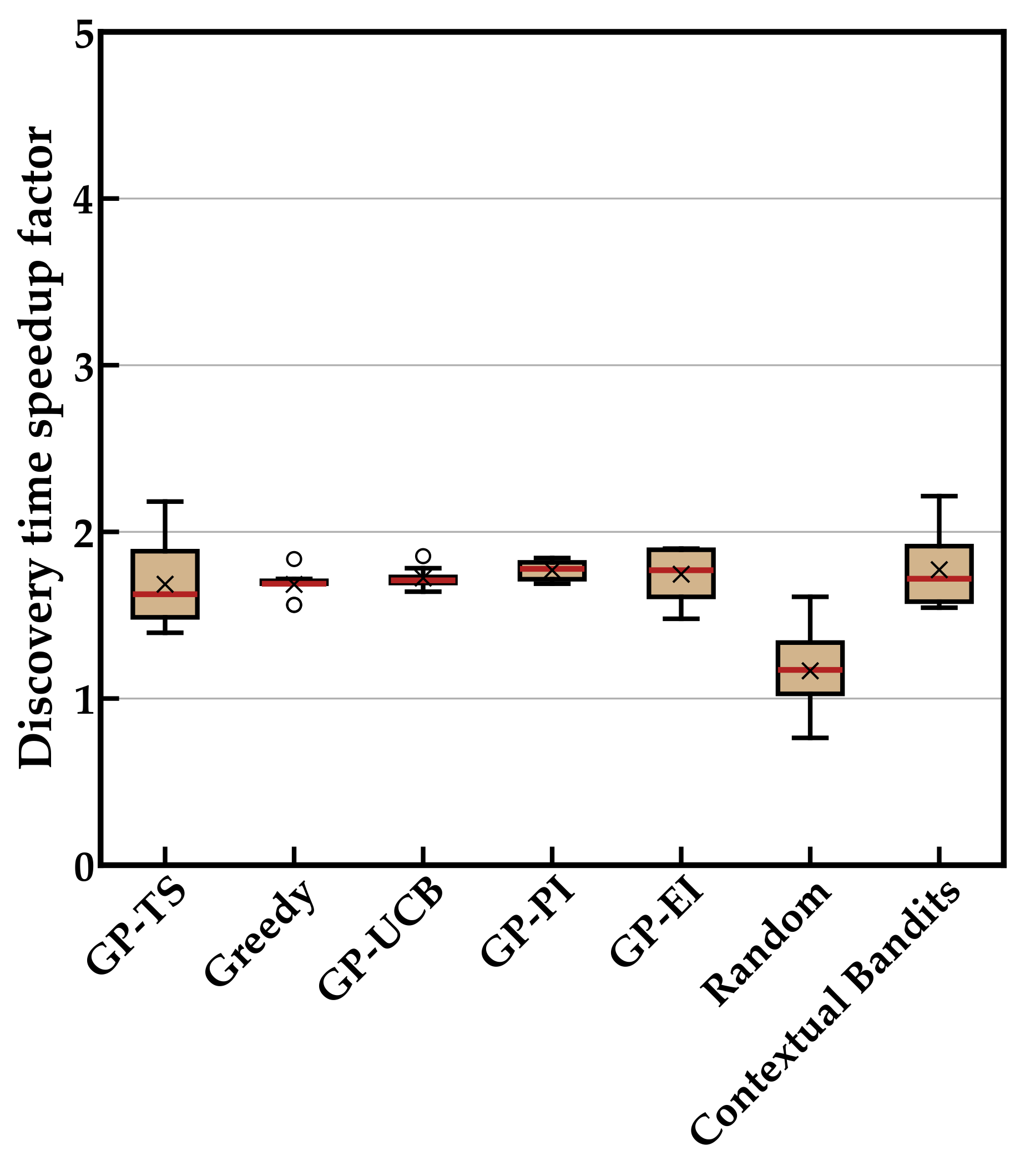}
		\end{minipage}
		\label{fig:speedup_FA}
	}
	\raggedright
	    \subfigure[]{
		\begin{minipage}{0.30\textwidth}
			\includegraphics[width=1\textwidth]{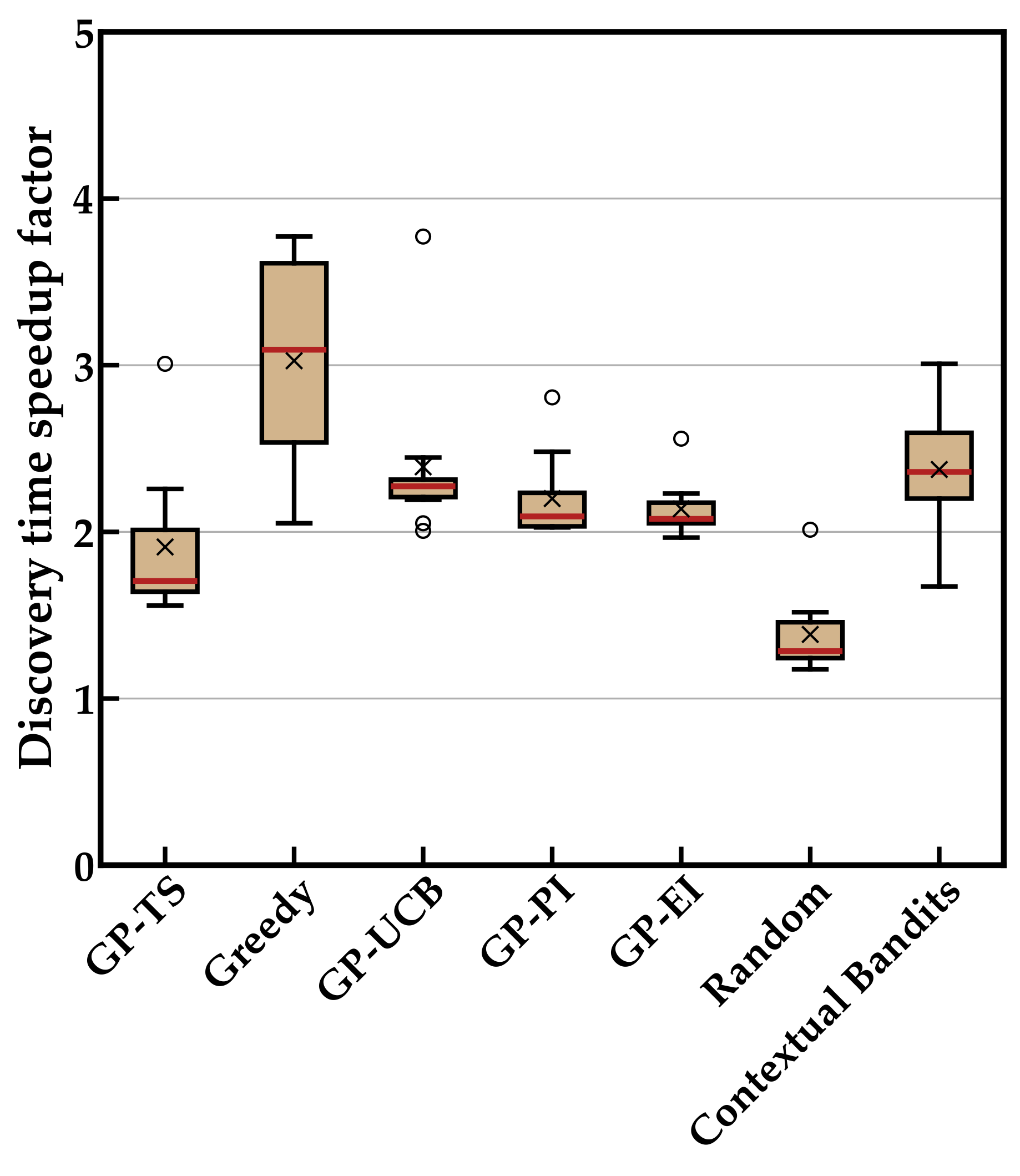}
		\end{minipage}
		\label{fig:speedup_NFA}
	}\hspace{2mm}
	
	\caption{Comparing the simulated path of material systems generated by data selection methods against the evolution of power conversion efficiency in the experimental literature. a) Both fullerene and non-fullerene acceptors included among candidate material systems b) fullerene acceptor only c) non-fullerene acceptors only. Fig. d-f shows the range of values obtained over \replaced{ten}{5} different starting material systems for each data selection method tested for the acceptors used in the row above. \added{In the box and whisker plot in Fig. d-f, the red line indicates the median of the data while the cross represents the mean of the data. Hollow circles are outliers.}}
	\label{fig:PSC_time_trend}
\end{figure}

The green line in Figure \ref{fig:PSC_time_trend}a-c shows how the power conversion efficiency of polymer solar cell systems reported in the literature has changed over time. Each data point corresponds to a single donor/acceptor system. In cases where a donor/acceptor system is reported multiple times in the literature, we take the median value and the timestamp for the paper corresponding to the median value \footnote{in cases where the number of data points $n$ is even, we use the PCE at index $\frac{n}{2}$ after the PCE values are sorted}. Observe that the value of PCE shows an upward trend till 2022 when a peak of 16.4 \% was obtained which is the highest value in \replaced{\Curated}{our data set}. 
Thus this plot captures a near complete picture of the evolution of the field of polymer solar cells. The discovery of new material systems has proceeded through trial and error which is why the PCE does not increase monotonically but fluctuates in value \added{over time}. The number of PCE values reported that under-perform the then-state-of-the-art is likely underestimated on this plot given that only material systems considered improvements get published. This results in a bias in the available data\cite{fujinuma2022big}.

We consider how this field may have evolved had it relied completely on data-driven methods to pick the next donor/acceptor combinations to evaluate rather than relying on trial and error. This allows us to empirically estimate how much faster we could have reached the end-point shown in Figure \ref{fig:PSC_time_trend}a-c.

Figure \ref{fig:workflow} shows the iterative process we used for material selection. The initial donor/acceptor system we used was OC$_{1}$C$_{10}$-PPV/PC$_{61}$BM \cite{dyakonov2004electrical} which was the earliest reported material system in \replaced{\Curated}{our data set}. \added{At each step, we trained a GPR model using all data already ``measured" and used that to make predictions on the remaining donor/acceptor pairs in \Curated.} We picked one donor/acceptor system that was the most ``promising" according to a few different possible strategies. We ``measured" the true power conversion efficiency by looking up the value from \replaced{\Curated}{our curated data set of known PCE values}. This new data point was then used to augment the training data set to train a new model and repeat the above cycle. The iterations were terminated when the material system with the highest value in our data set was identified. This is identical to active learning with the key difference being that we can compare active learning generated paths \added{of donor/acceptor pairs} against how the field evolved and estimate the speedup. We make three key assumptions in this analysis:

\begin{enumerate}
    \item The list of donor/acceptor systems used as candidate material systems is the full list of material systems reported in \replaced{\Curated}{our curated data set}. While we do not assume fore-knowledge of the actual value of PCE, coming up with viable new material systems was part of the evolution of this field which we are unable to realistically simulate. A fairer analysis would assume a much larger candidate list of donor/acceptor systems which would be created without the bias \added{that they were reported in the polymer solar cell literature}. There would however be no way to ``measure" the value of PCE for these material systems with the same fidelity as experimentally reported values. 
    \item The timestamp associated with each new donor/acceptor system that is picked is the same as the timestamp for an entry of the corresponding index in the \replaced{experimental}{ground truth} data. As the experimentally reported results come from many different research groups, this assumption is reasonable if we assume that many different research groups are involved in this data-driven workflow. It may even be possible for funding agencies to allocate resources more efficiently if the iterative methodology described in this section were to be followed. 
    \item \added{Our simplified problem formulation considers only the discovery of donor/acceptor pairs which are the most critical component of a polymer solar cell and does not consider processing/fabrication.} 
\end{enumerate}

As we see in Figure \ref{fig:PSC_time_trend}a-f, \replaced{the}{all} data selection methods used in this work outperform trial and error \added{and random data selection}. We compared different methods using the discovery time speedup factor (DTSF) which is defined below:

\begin{equation}
    \text{DTSF} = \frac{t^{exp}_{max}-t_0}{t^{sim}_{max}-t_0}
\end{equation}

where $t_0$ is the timestamp associated with the earliest data point in \replaced{\Curated}{our data set}, $t^{exp}_{max}$ is the timestamp for the paper when the highest experimental PCE was reported in \replaced{\Curated}{our data set}, $t^{sim}_{max}$ is the timestamp corresponding to the index at which the simulated path reaches the same maximum PCE. The timestamp here is the date converted to a rational number \replaced{in which}{wherein} the fractional part is the day and month converted to a fraction of a full year and the integer part is the year.

Intuitively this gives a sense of how much faster machine learning methods can reach the same end-point as trial and error methods. When fullerenes, as well as non-fullerenes, are used as \added{candidate} acceptors, we obtain the greatest speedup of a factor of about 4 while the speedup factor is lower when only one of fullerenes or non-fullerenes are used as \added{candidate} acceptors, reducing to a factor of $\sim$2. In the context of polymer solar cell donor/acceptor discovery, this translates to a time saving of $\sim$15 years. This is one of the first quantitative estimates of the time saved by using machine learning for materials discovery that we are aware of \added{using historical data of how materials discovery evolved}. Figure \ref{fig:PSC_time_trend}a-c assume a fixed starting donor/acceptor system, namely OC$_{1}$C$_{10}$-PPV/PC$_{61}$BM. For all methods tested except GP-TS and linear contextual bandits, a fixed starting point would result in a fixed path of donor/acceptor pairs being selected as a GPR model will always generate the same predictions given fixed training data and fixed hyperparameters. In the case of GP-TS and linear contextual bandits, however, there is a sampling step involved in picking a new material which inherently introduces randomness. We used the first \replaced{ten}{five} donor/acceptor pairs reported in \replaced{\Curated}{our curated data set} to estimate the range of speedups obtained for each method. These \replaced{ten}{five} donor/acceptor systems belong to the same generation of polymer solar cells and are listed in Supporting Information Section 3. When error bars are accounted for, we notice, from Figure \ref{fig:PSC_time_trend}d-f that \replaced{GP-UCB, GP-EI, GP-PI and contextual bandits perform similarly.}{there is no statistically significant difference between the different methods that we tested.} \added{GP-TS appears to underperform the other data selection methods in Figer \ref{fig:PSC_time_trend}d and f while Greedy outperforms the other data selection methods in Figure \ref{fig:PSC_time_trend}f.} \replaced{Since Greedy}{GP-EI} has a higher value on average\replaced{,}{and so} we plot the path generated by \replaced{Greedy}{GP-EI} in Figure \ref{fig:PSC_time_trend}a-c. \added{GP-UCB has the tightest variance across all data selection methods tested, for all acceptor types. This indicates that trading off exploration and exploitation is a sound strategy to be robust to the starting material system of the active learning cycle.} Note that linear contextual bandits which is a linear method and therefore much faster than non-parametric methods like GPR, has a similar speedup factor as GPR-based methods. This indicates that the predictive capabilities of a model have little to do with how quickly it can identify the best-performing material system. We shall explore this more fully in the next section.




\subsubsection{Analyzing the predictions from data selection methods}

\begin{figure}[!h]
	\raggedleft
	\subfigure[]{
		\begin{minipage}{0.30\textwidth}
			\includegraphics[width=1\textwidth]{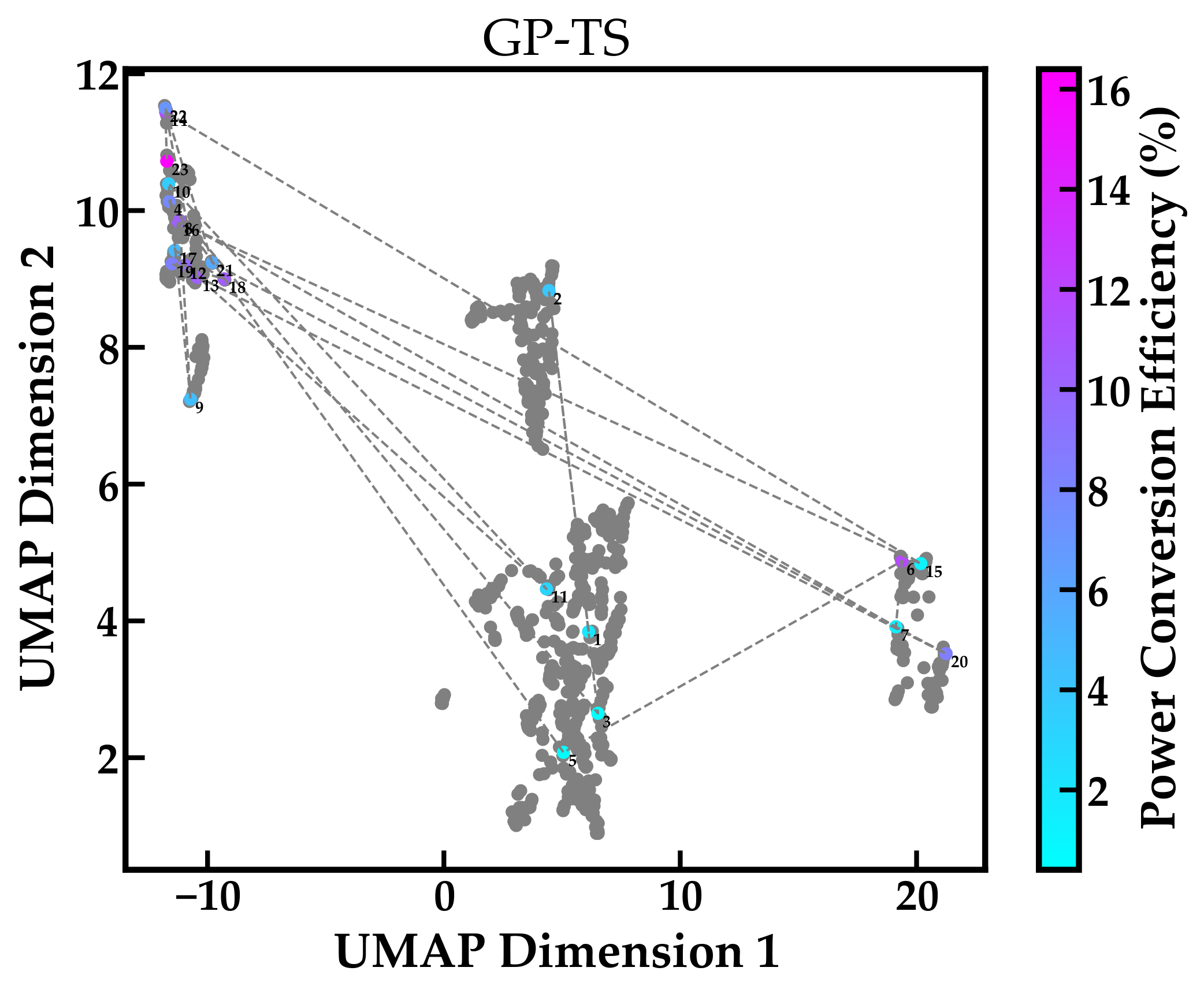}
		\end{minipage}
		\label{fig:path_GP_TS}
	}\hspace{2mm}
	\centering
	\subfigure[]{
		\begin{minipage}{0.30\textwidth}
			\includegraphics[width=1\textwidth]{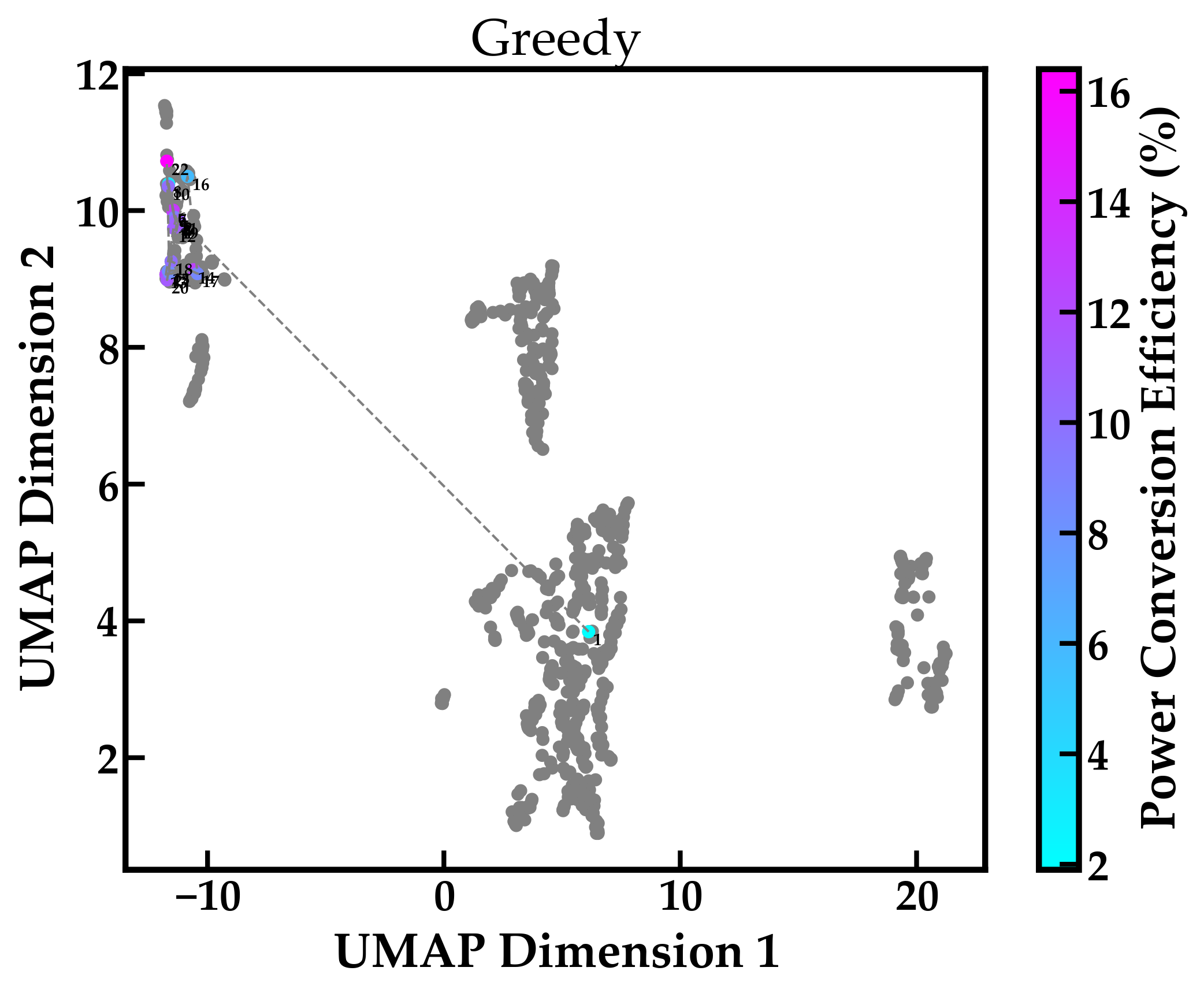}
		\end{minipage}
		\label{fig:path_greedy}
	}
	\raggedright
	    \subfigure[]{
		\begin{minipage}{0.30\textwidth}
			\includegraphics[width=1\textwidth]{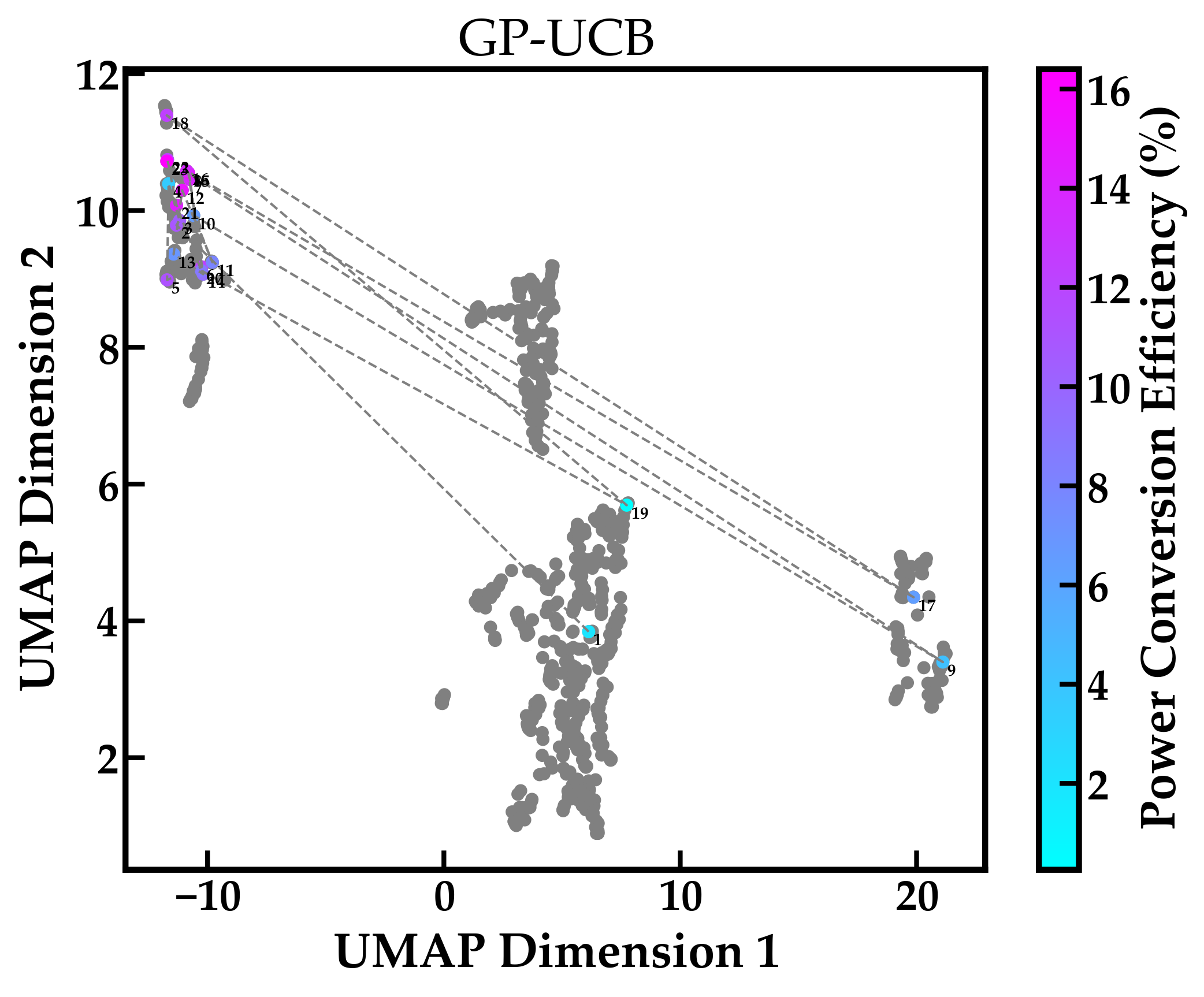}
		\end{minipage}
		\label{fig:path_GP_UCB}
	}\hspace{2mm}
	
	\raggedleft
	\subfigure[]{
		\begin{minipage}{0.30\textwidth}
			\includegraphics[width=1\textwidth]{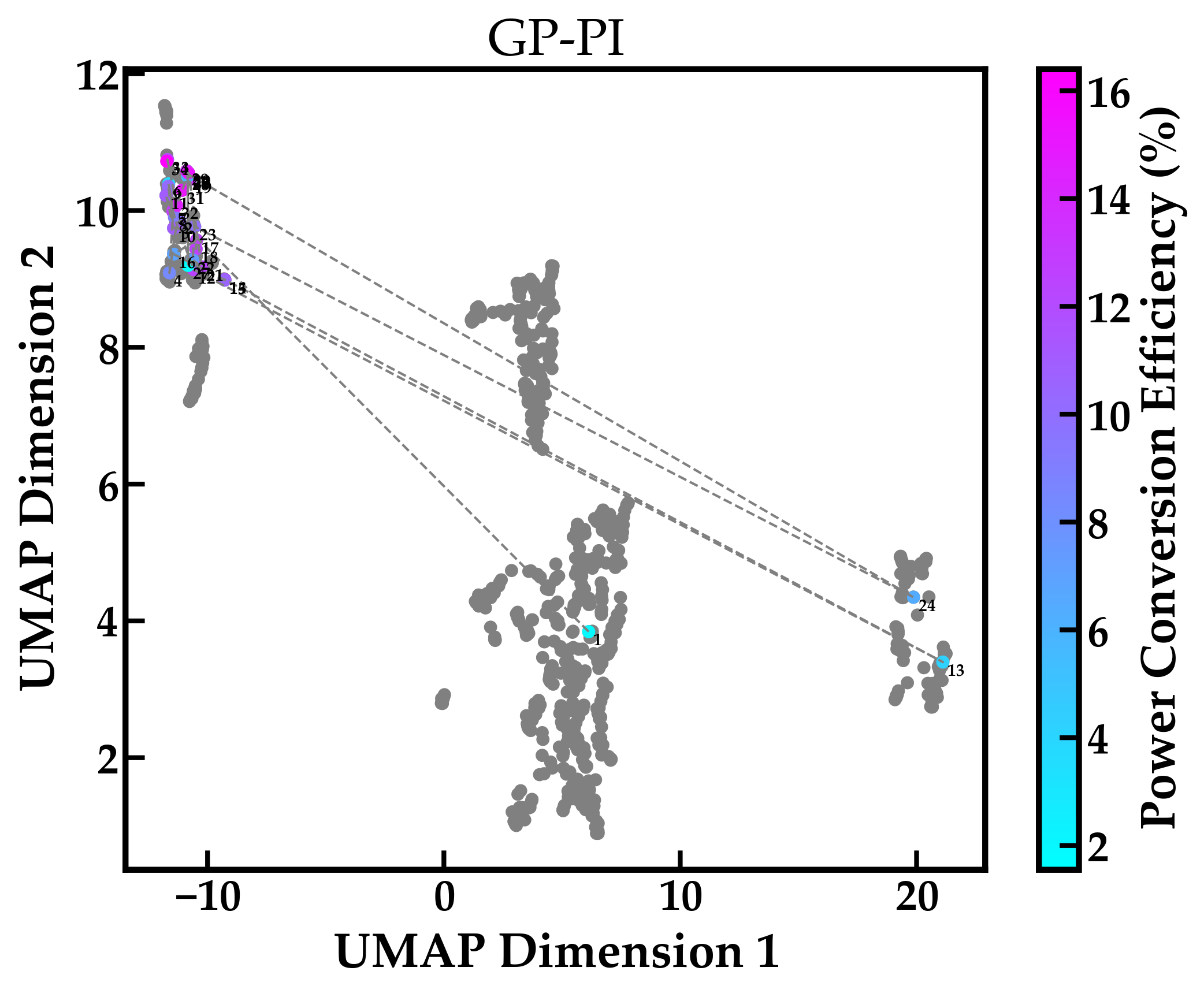}
		\end{minipage}
		\label{fig:path_GP_PI}
	}\hspace{2mm}
	\centering
    \subfigure[]{
		\begin{minipage}{0.30\textwidth}
			\includegraphics[width=1\textwidth]{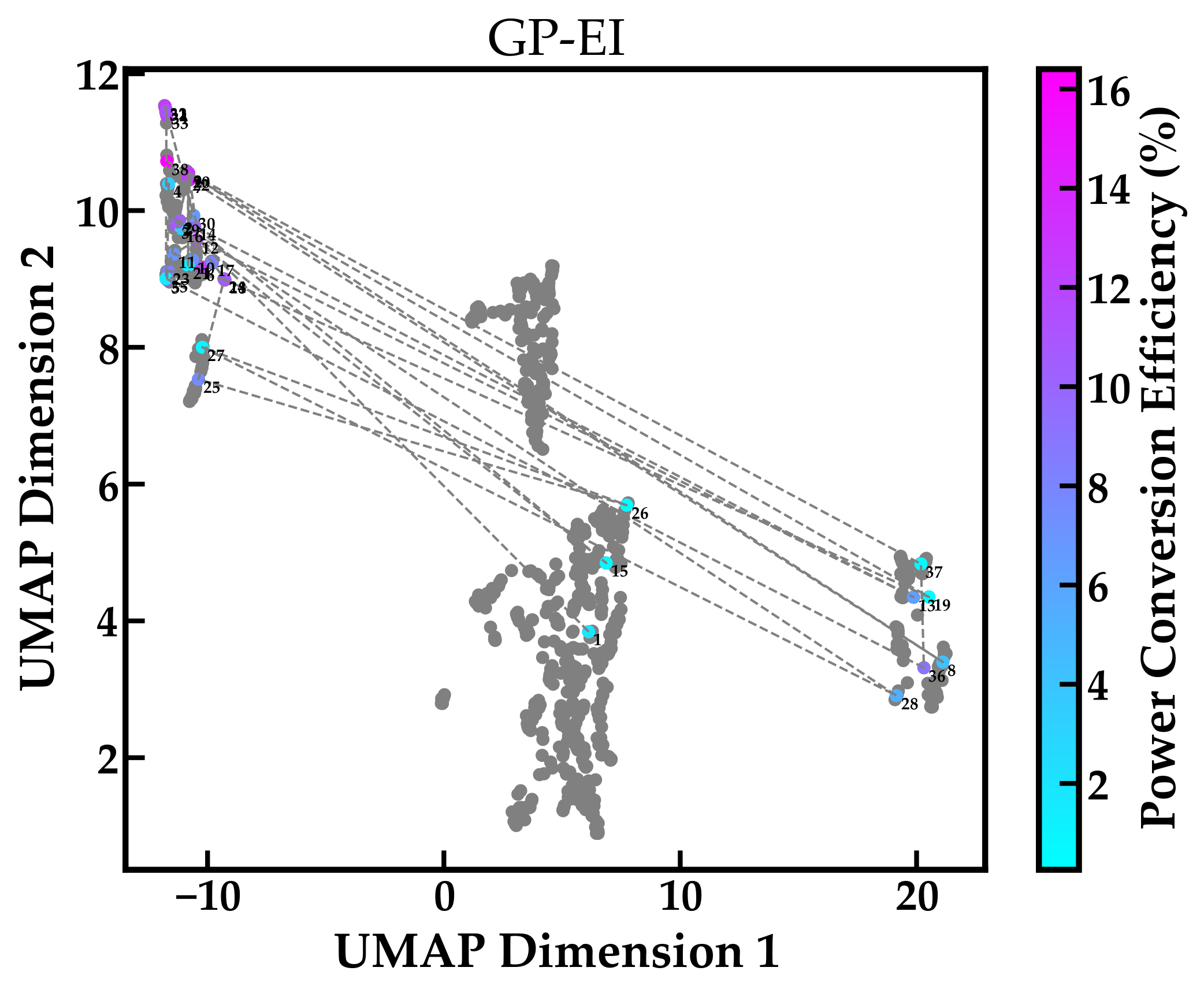}
		\end{minipage}
		\label{fig:path_GP_EI}
	}
	\raggedright
	\subfigure[]{
		\begin{minipage}{0.30\textwidth}
			\includegraphics[width=1\textwidth]{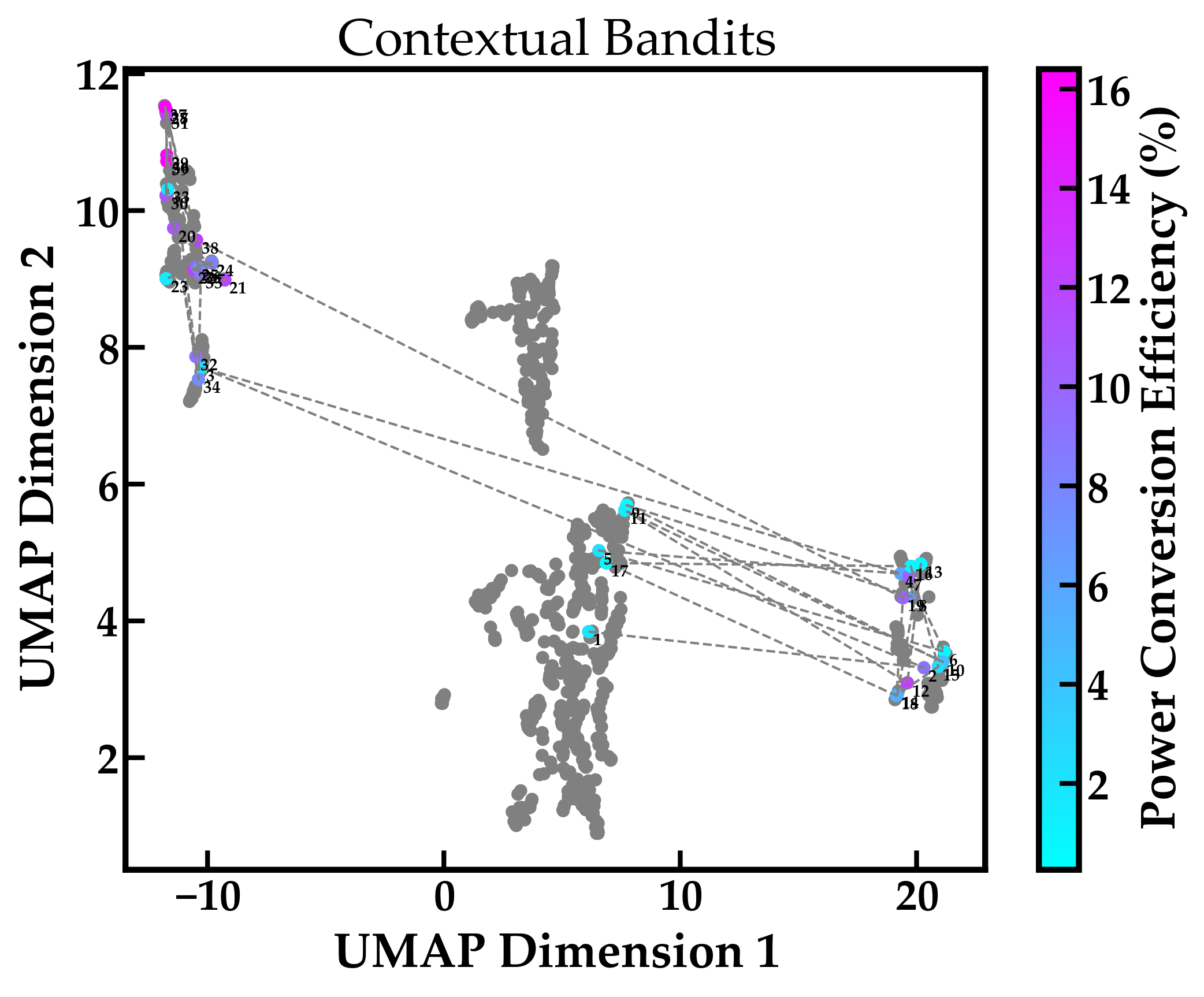}
		\end{minipage}
		\label{fig:path_GP_LCB}
	}\hspace{2mm}
	
	\caption{The simulated path of material systems for each data selection method with the material system vectors being embedded in two dimensions using UMAP. The left most cluster in each figure corresponds to non-fullerene small molecule acceptors. The middle two clusters correspond to fullerene acceptors within which the top cluster corresponds to donors with fused aromatic rings and the bottom cluster corresponds to all other donors. The rightmost cluster consists of polymer acceptors. The plots correspond to a) GP-TS b) Greedy c) GP-UCB d) GP-PI e) GP-EI f) Linear contextual bandits}
	\label{fig:PSC_path_both}
\end{figure}

This section examines the path of donor/acceptor pairs picked by various data selection strategies. We projected the vector representations of the concatenated donor/acceptor material fingerprint to two dimensions using Uniform Manifold Approximation and Projection (UMAP) \cite{mcinnes2018umap}. We considered fullerene as well as non-fullerene acceptors. This projection results in four clusters forming (Figure \ref{fig:PSC_path_both}). The leftmost cluster corresponds to non-fullerene small molecule acceptors and the remaining clusters are described in the caption. The paths shown in Figure \ref{fig:PSC_path_both} were generated using OC$_{1}$C$_{10}$-PPV/PC$_{61}$BM as the initial donor/acceptor pair.
Note from Figure \ref{fig:PSC_path_both}b \replaced{}{and e} that \replaced{greedy}{both greedy and GP-EI} jumps to the non-fullerene small molecule space after the first iteration and stays there during all subsequent iterations. This space is known to contain the candidates most likely to have high power conversion efficiency. This is likely why greedy has higher average speedups compared to other methods as seen in Figure \ref{fig:PSC_time_trend}a and c. All other methods spend at least one or more iterations in exploring other parts of the donor/acceptor space. Observe also that greedy \replaced{}{and GP-EI} when applied to non-fullerene acceptors and both fullerene and non-fullerene acceptors differ in the first material system only. All subsequent iterations for both cases are restricted to the space of non-fullerenes. Yet, there is a statistically significant difference in the time taken to find the optimal material system in both cases (refer Figure \ref{fig:PSC_time_trend}\added{d and f}) with the non-fullerene only case taking significantly longer. We posit that the initial inclusion of the fullerene acceptor or more generally the inclusion of some initial material system that is different from the space where the optimal material system is expected to lie, would accelerate convergence to the optimal material system. This could be because the initial diversity helps give the model a better ``view" of the \replaced{}{overall} material space compared to if the candidates were less diverse. The latter can at best present the model with a very local region of the material space, thus requiring \replaced{}{the model to spend} more iterations to understand the best direction to traverse in material space.

\begin{figure}[!h]
	\raggedleft
	\subfigure[]{
		\begin{minipage}{0.30\textwidth}
			\includegraphics[width=1\textwidth]{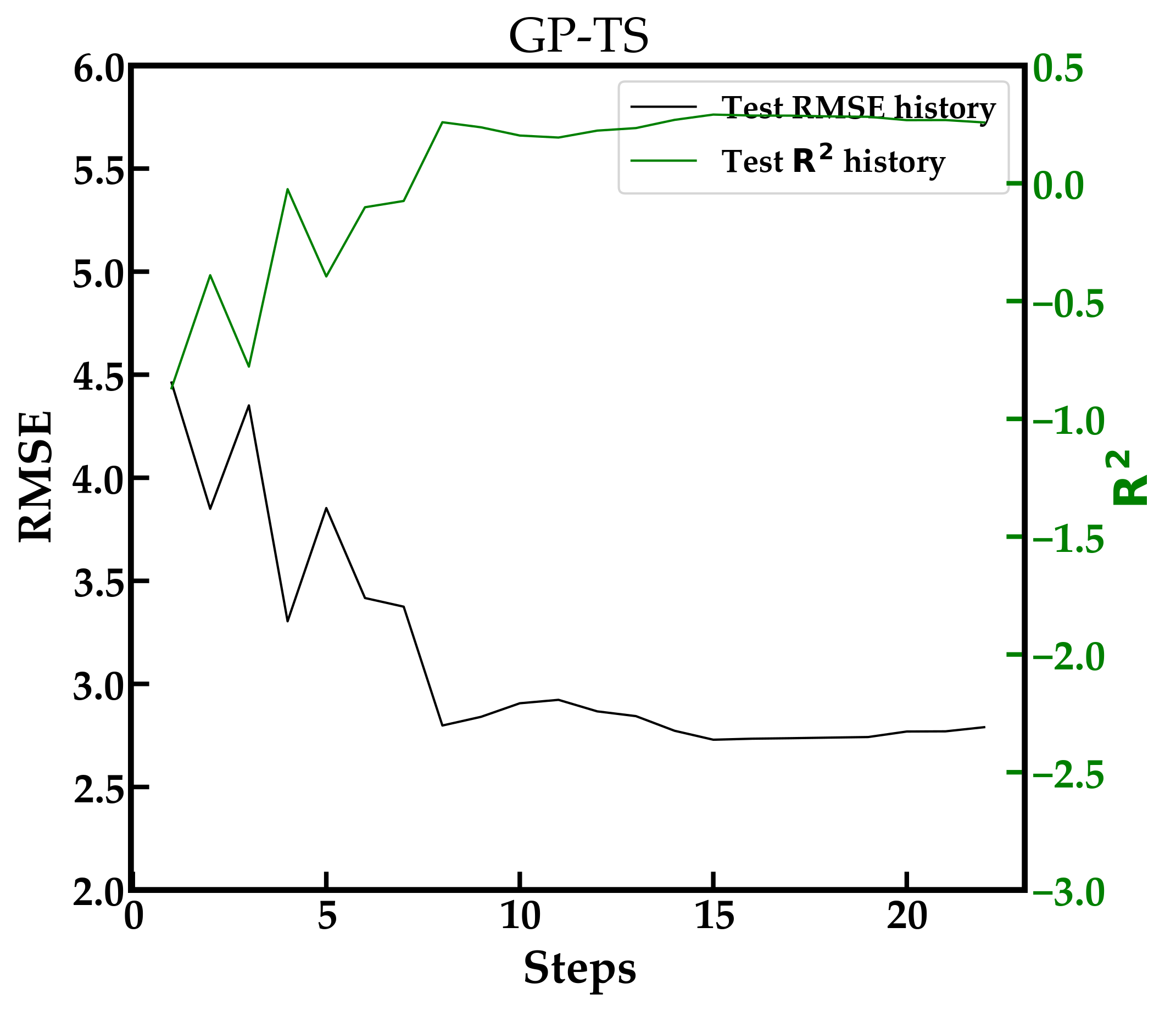}
		\end{minipage}
		\label{fig:metrics_GP_TS}
	}\hspace{2mm}
	\centering
	\subfigure[]{
		\begin{minipage}{0.30\textwidth}
			\includegraphics[width=1\textwidth]{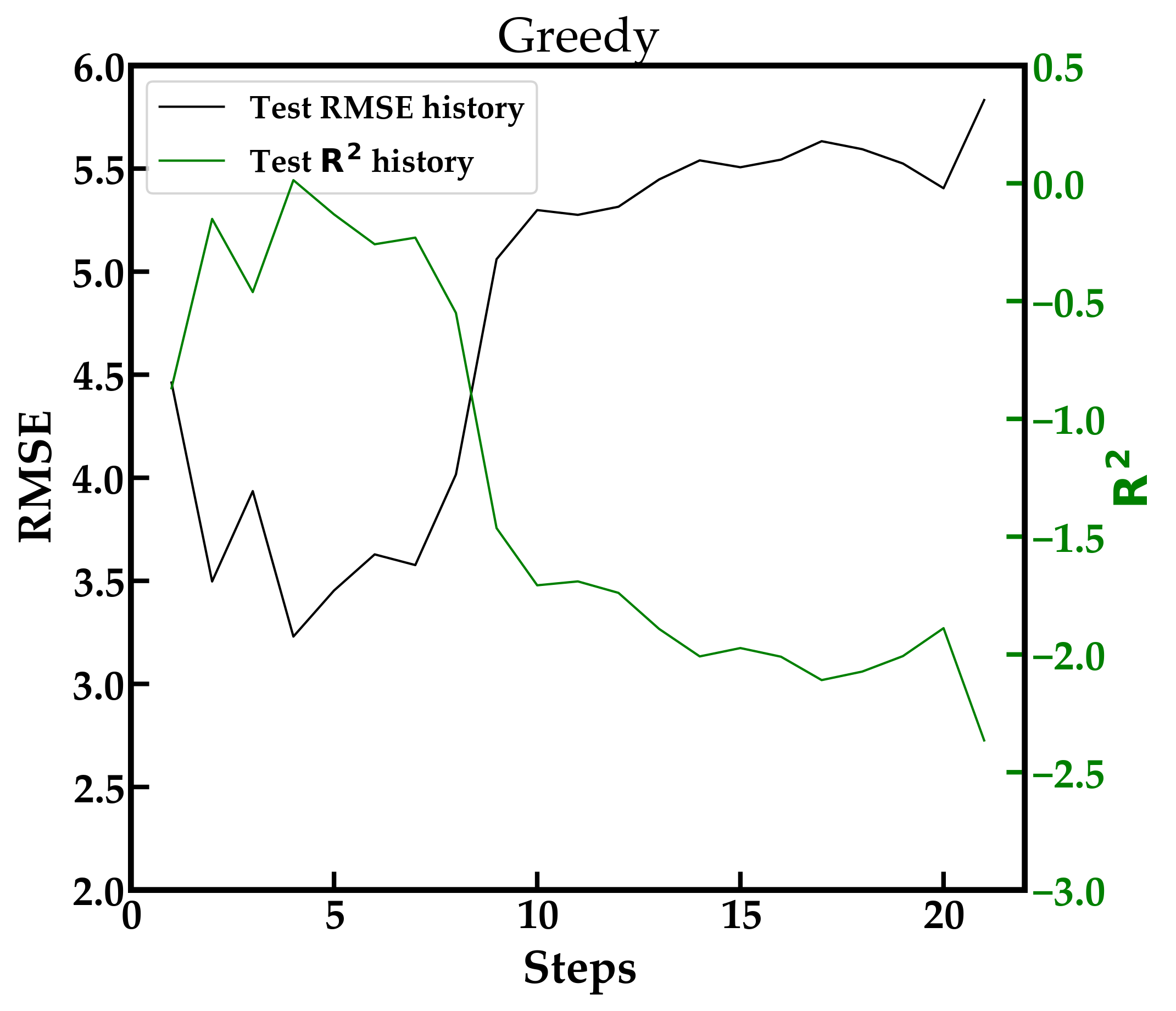}
		\end{minipage}
		\label{fig:metrics_greedy}
	}
	\raggedright
	    \subfigure[]{
		\begin{minipage}{0.30\textwidth}
			\includegraphics[width=1\textwidth]{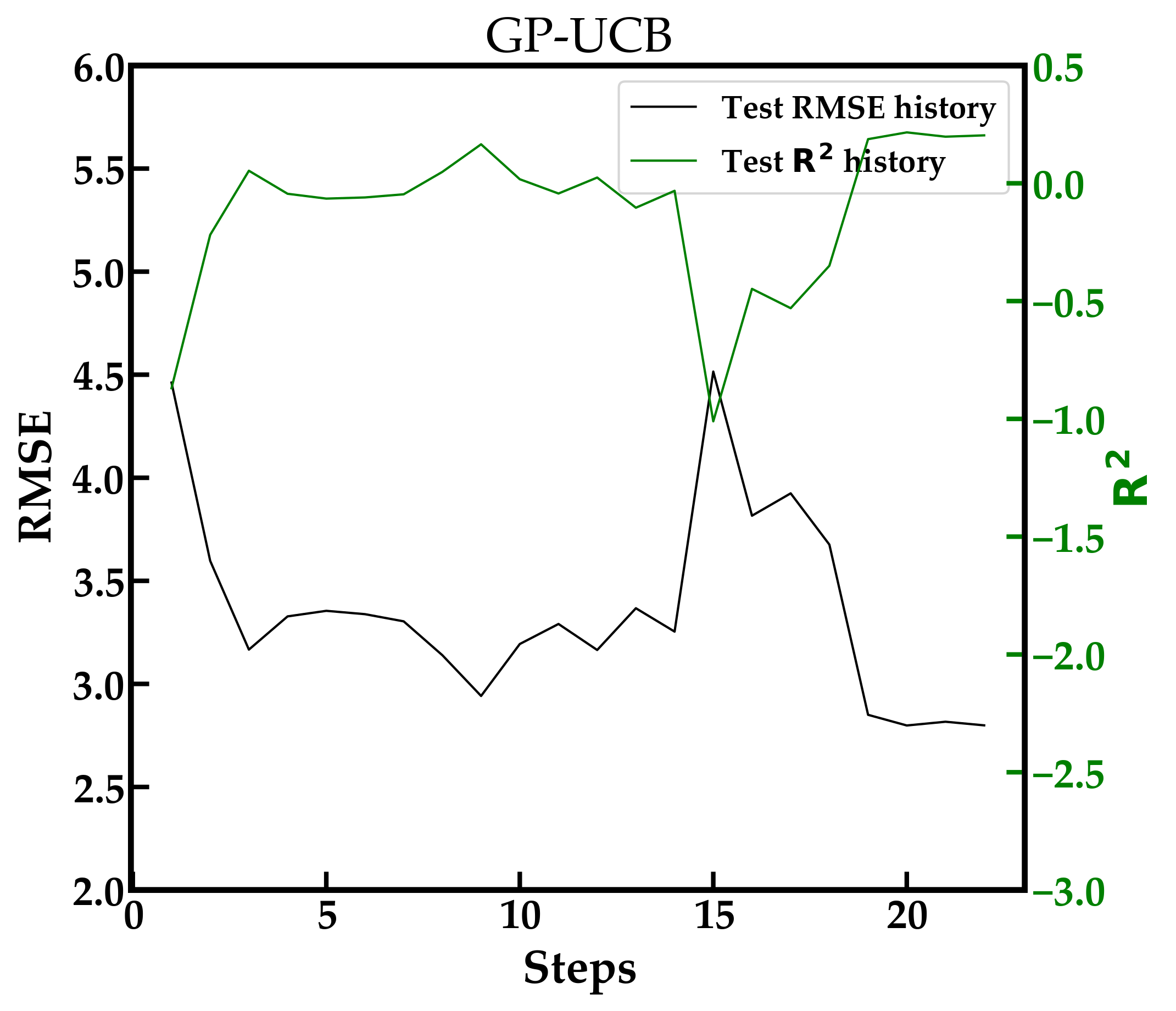}
		\end{minipage}
		\label{fig:metrics_GP_UCB}
	}\hspace{2mm}
	
	\raggedleft
	\subfigure[]{
		\begin{minipage}{0.30\textwidth}
			\includegraphics[width=1\textwidth]{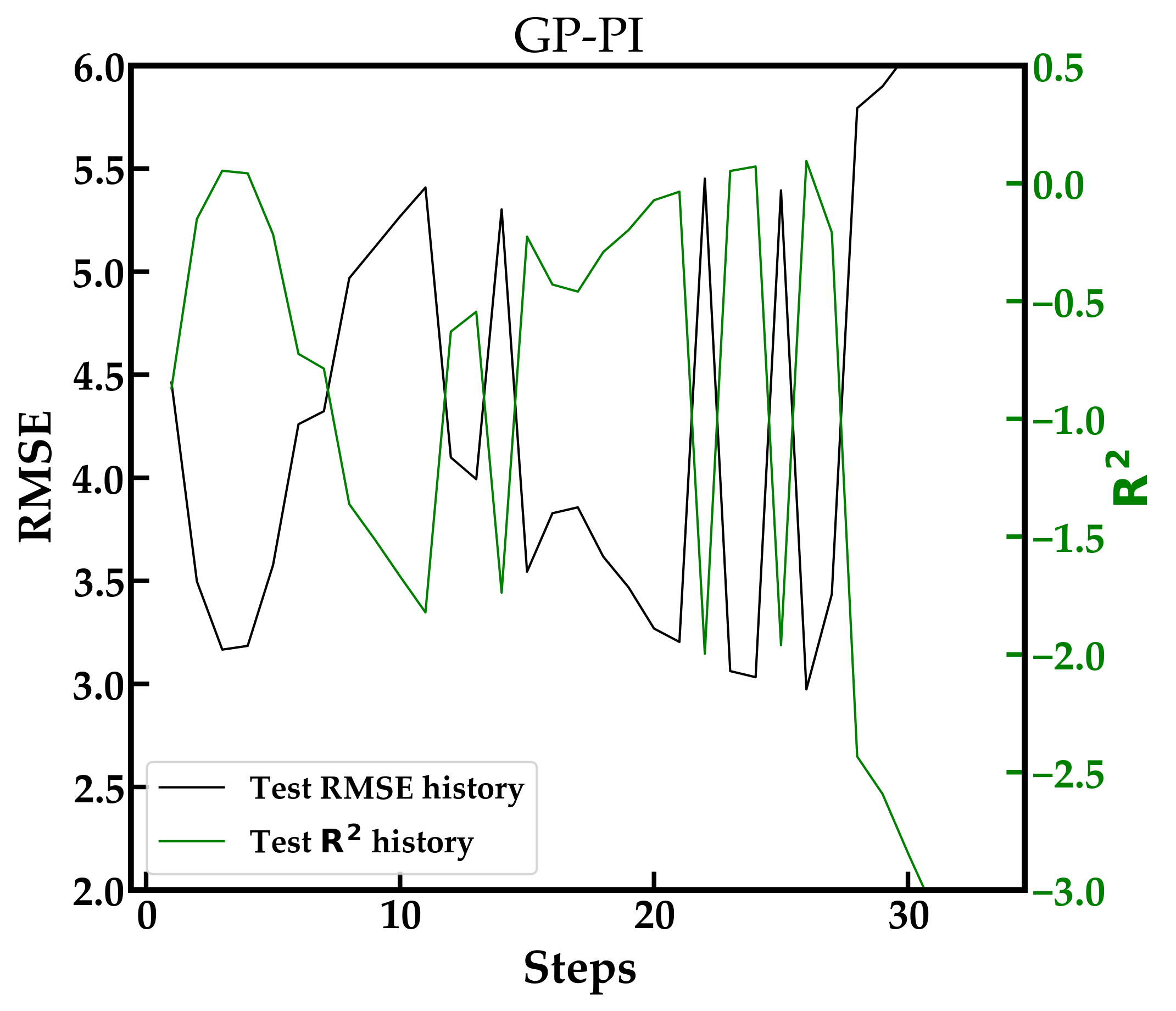}
		\end{minipage}
		\label{fig:metrics_GP_PI}
	}\hspace{2mm}
	\centering
    \subfigure[]{
		\begin{minipage}{0.30\textwidth}
			\includegraphics[width=1\textwidth]{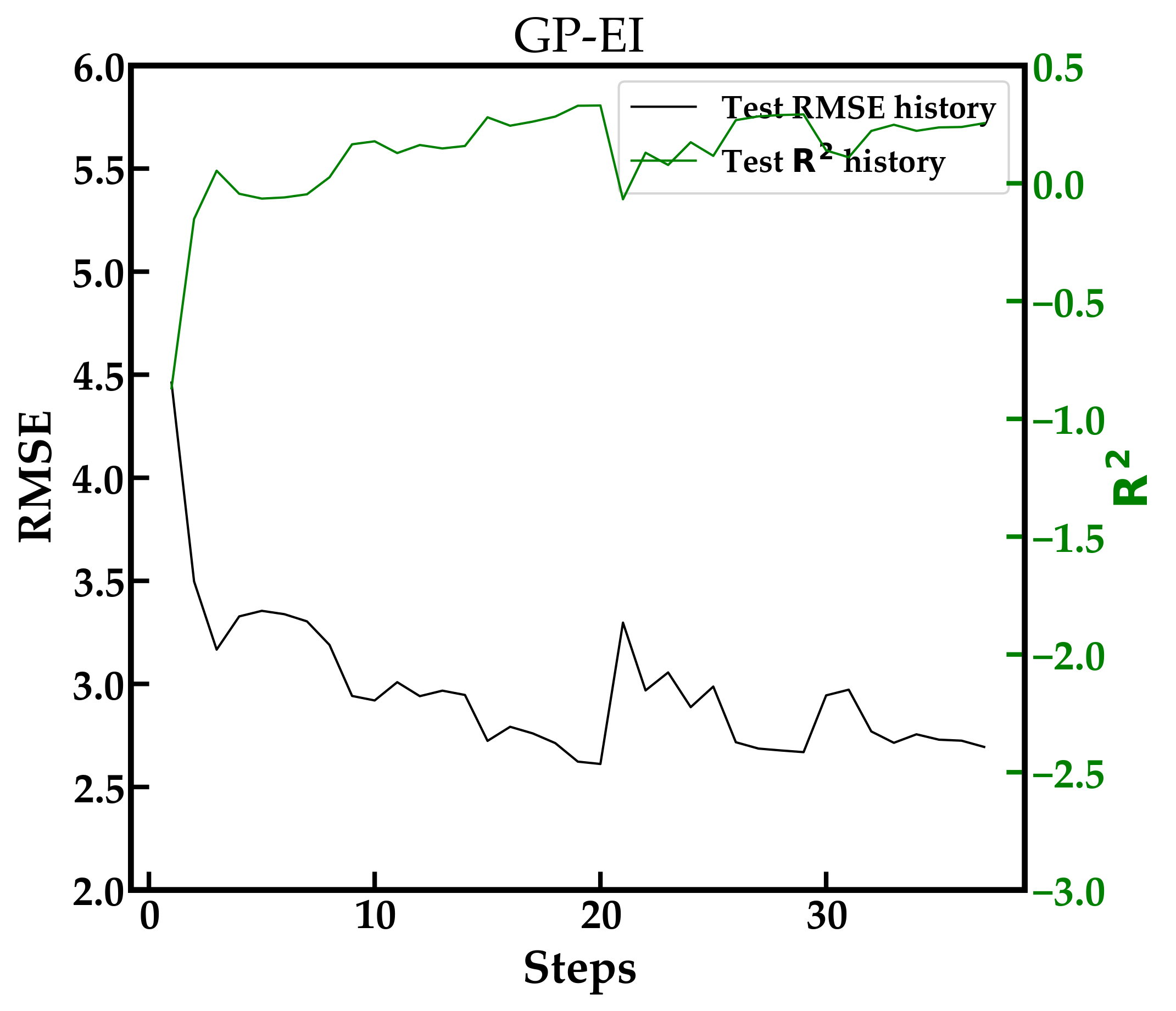}
		\end{minipage}
		\label{fig:metrics_GP_EI}
	}
	
	\caption{Evolution of the predictive performance of the models trained using different data selection methods using OC$_{1}$C$_{10}$-PPV/PC$_{61}$BM as the initial donor/acceptor pair. The data selection methods compared are a) GP-TS b) Greedy c) GP-UCB d) GP-PI e) GP-EI}
	\label{fig:PSC_predictor_path_both}
\end{figure}

At each iteration, we measured the predictive performance of the trained GPR model by comparing its predictions against all points not yet added to the test set (Figure \ref{fig:PSC_predictor_path_both}). The data points used for measuring test performance are reduced by one point at every iteration and have some points that are different for each selection method. However, the number of points selected during a typical active learning run (20-30) is small compared to the total number of data points (835) which makes this a reasonable approach to infer the general trend in the predictive capabilities of the models being trained. Note that there is no consistent improvement in the predictive performance of the models being trained as the number of data points in the training set increases, for most of the selection methods we test. This is consistent with results reported in Ref. \citenum{borg2023quantifying}. The notable exception to this however is GP-TS in which successive models have improved predictive performance. This intuitively makes sense as all other methods pick material systems by directly optimizing for high power conversion efficiency and therefore bias the trained model using the picked data points. GP-TS on the other hand picks material systems by maximizing over PCEs sampled from the predicted distribution for each material system and thus samples a more diverse set of data points as seen in Figure \ref{fig:PSC_path_both}a. \added{This diverse sampling is likely what leads to GP-TS being slower than other data selection methods as seen in Figure \ref{fig:PSC_time_trend}d-f. Thus this data selection strategy allows trading-off speed of discovery for training a stronger predictive model.} By design, GP-TS reaches the optimal material state quickly and in parallel trains models with strong predictive performance and is the only method out of the methods we test that \added{consistently} achieves this balance \added{across acceptor types}. The trend observed in Figure \ref{fig:PSC_predictor_path_both} is reported for both types of acceptors. The same trend for GP-TS is observed for iterations performed over fullerene acceptors and non-fullerene acceptor \added{candidates}, reported in Supporting Information Section 4.

\section{Summary and Outlook}

A pipeline that extracts material property data from published literature which is then used for training property prediction models \replaced{}{and in turn generates novel insights} was built. We found promising donor/acceptor combinations that have not yet been reported in the literature, by training models to predict power conversion efficiency. We used the timestamp associated with our literature-extracted data set to demonstrate that data-driven methods alone would have discovered the most promising donor/acceptor system in only one-fourth of the time it took through trial and error. This allows for a stronger case to be made to policymakers to encourage further use of ML methods for materials discovery.
\replaced{}{We also discover that Gaussian Process-Thompson Sampling as an active learning strategy discovers the optimal donor/acceptor system as quickly as any of the other strategies tested but also simultaneously results in training a strong property predictor, unlike other strategies.}
\added{We summarize some best practices learned from this study for deploying active learning for materials discovery}

\begin{enumerate}
    \item \added{Gaussian Process-Thompson sampling should be used as the active learning strategy if training a strong property predictor is desired.}
    \item \added{If robustness to the starting material is desired, then GP-UCB is the recommended strategy which will also on average lead to fast discovery of the optimal material system.}
    \item \added{Using a diverse initial set of candidates during the active learning cycle helps the model learn the optimization landscape more quickly and is empirically observed to lead to faster convergence.}
\end{enumerate}
There are some key limitations of our study that we highlight below.

\begin{enumerate}
    \item The list of candidate material systems we pick from, is the set of donor/acceptor systems that have already been tested in the literature. The structures that were tested much later in the literature could only be discovered due to the experimental trial and error that preceded it.
    \item We assume that the time intervals of measurement in the data-driven view are identical to the trial and error approach, i.e., new materials through data-driven approaches can be discovered at the same rate at which experimental papers containing new donor/acceptor systems were published.
    \item \added{Our simplified formulation of the polymer solar cell optimization problem does not consider other aspects of the cell such as the electrodes, electron/hole transport layer, and fabrication/processing of the cell. These factors would affect our estimate of the speedup of ML methods over trial-and-error experiments.}
\end{enumerate}

This study views active learning as a sequential process whereby a material is ``chosen" by the model and tested by the scientist. The only input from the scientist is the list of materials provided initially. The difficulty of coming up with a perfect initial list ensures, however, that scientists and machines would need to coordinate more closely in practice.
The insights generated by the selection process can guide scientific intuition to update the list of materials continuously. 
An example of this is provided in Supporting Information Section 5.

One of the bottlenecks when going from NLP-extracted data to a data set usable for training ML models is obtaining SMILES strings for the donor and acceptor. This had to be done manually during data curation. There are computer vision tools available that can convert structures to SMILES strings such as MolScribe \cite{molscribe} and OSRA \cite{osra}. However, donor or acceptor structures found in the literature are typically part of a larger figure such as a reaction scheme. Robust segmentation of relevant structures combined with conversion of structures to SMILES strings is necessary to bridge the gap between NLP extracted data and trained property predictors. Furthermore, the conversion of polymer structures to SMILES strings is not \added{yet} possible through these tools\replaced{}{and is still an open question}.



We also limited our focus to optimizing a single material property, namely the PCE. In practice, however, materials scientists are often interested in optimizing multiple properties simultaneously which may be inversely co-related. In Ref. \citenum{khatamsaz2022multi}, for instance, the authors find optimal multi-principal element alloys while optimizing two ductility indicators (Pugh’s Ratio and Cauchy pressure). Multi-objective Bayesian approaches using hypervolume indicators are a promising approach for solving such problems \cite{yang2007novel, emmerich2011hypervolume}. In organic photovoltaics, a possible direction of inquiry would be a simultaneous optimization of the efficiency, flexibility, and stability of the solar cell. PSCs need to be stable and retain their power conversion efficiency with usage. This is typically inversely co-related with PCE. Higher flexibility PSCs usually have weaker mechanical properties and therefore lower PCE \cite{xu2018thermally, gevorgyan2017improving, rafique2018fundamentals}.

The ability to scale up the pipeline we have built to other properties and materials classes will significantly speed up the development and deployment of new materials. This in turn will accelerate scientific discovery across many critical applications such as clean energy, healthcare, and devices.



\section{Supporting Information}
\added{Additional experimental details, results, and theoretical background on methods used}

\section{Acknowledgements}
This work was also supported by the Office of Naval Research through grants N00014-19-1-2103 and N00014-20-1-2175. Pranav Shetty was funded by a fellowship by JPMorgan Chase \& Co. that helped to support this research. Any views or opinions expressed herein are solely those of the authors listed, and may differ from the views and opinions expressed by JPMorgan Chase \& Co. or its affiliates.

\section{Data and software availability}
All code and data needed to reproduce the results in this paper can be found at \url{https://github.com/pranav-s/PolymerSolarCellsML}

\bibliography{bibliography_new}

\providecommand{\latin}[1]{#1}
\makeatletter
\providecommand{\doi}
  {\begingroup\let\do\@makeother\dospecials
  \catcode`\{=1 \catcode`\}=2 \doi@aux}
\providecommand{\doi@aux}[1]{\endgroup\texttt{#1}}
\makeatother
\providecommand*\mcitethebibliography{\thebibliography}
\csname @ifundefined\endcsname{endmcitethebibliography}  {\let\endmcitethebibliography\endthebibliography}{}
\begin{mcitethebibliography}{76}
\providecommand*\natexlab[1]{#1}
\providecommand*\mciteSetBstSublistMode[1]{}
\providecommand*\mciteSetBstMaxWidthForm[2]{}
\providecommand*\mciteBstWouldAddEndPuncttrue
  {\def\EndOfBibitem{\unskip.}}
\providecommand*\mciteBstWouldAddEndPunctfalse
  {\let\EndOfBibitem\relax}
\providecommand*\mciteSetBstMidEndSepPunct[3]{}
\providecommand*\mciteSetBstSublistLabelBeginEnd[3]{}
\providecommand*\EndOfBibitem{}
\mciteSetBstSublistMode{f}
\mciteSetBstMaxWidthForm{subitem}{(\alph{mcitesubitemcount})}
\mciteSetBstSublistLabelBeginEnd
  {\mcitemaxwidthsubitemform\space}
  {\relax}
  {\relax}

\bibitem[Mannodi-Kanakkithodi \latin{et~al.}(2018)Mannodi-Kanakkithodi, Chandrasekaran, Kim, Huan, Pilania, Botu, and Ramprasad]{mannodi2018scoping}
Mannodi-Kanakkithodi,~A.; Chandrasekaran,~A.; Kim,~C.; Huan,~T.~D.; Pilania,~G.; Botu,~V.; Ramprasad,~R. Scoping the polymer genome: A roadmap for rational polymer dielectrics design and beyond. \emph{Materials Today} \textbf{2018}, \emph{21}, 785--796\relax
\mciteBstWouldAddEndPuncttrue
\mciteSetBstMidEndSepPunct{\mcitedefaultmidpunct}
{\mcitedefaultendpunct}{\mcitedefaultseppunct}\relax
\EndOfBibitem
\bibitem[Ma \latin{et~al.}(2015)Ma, Sharma, Baldwin, Tefferi, Offenbach, Cakmak, Weiss, Cao, Ramprasad, and Sotzing]{ma2015rational}
Ma,~R.; Sharma,~V.; Baldwin,~A.~F.; Tefferi,~M.; Offenbach,~I.; Cakmak,~M.; Weiss,~R.; Cao,~Y.; Ramprasad,~R.; Sotzing,~G.~A. Rational design and synthesis of polythioureas as capacitor dielectrics. \emph{Journal of Materials Chemistry A} \textbf{2015}, \emph{3}, 14845--14852\relax
\mciteBstWouldAddEndPuncttrue
\mciteSetBstMidEndSepPunct{\mcitedefaultmidpunct}
{\mcitedefaultendpunct}{\mcitedefaultseppunct}\relax
\EndOfBibitem
\bibitem[Wu \latin{et~al.}(2021)Wu, Chen, Deshmukh, Kamal, Li, Shetty, Zhou, Sahu, Tran, Sotzing, Ramprasad, and Cao]{wu2021dielectric}
Wu,~C.; Chen,~L.; Deshmukh,~A.; Kamal,~D.; Li,~Z.; Shetty,~P.; Zhou,~J.; Sahu,~H.; Tran,~H.; Sotzing,~G.; Ramprasad,~R.; Cao,~Y. Dielectric polymers tolerant to electric field and temperature extremes: Integration of phenomenology, informatics, and experimental validation. \emph{ACS Applied Materials \& Interfaces} \textbf{2021}, \emph{13}, 53416--53424\relax
\mciteBstWouldAddEndPuncttrue
\mciteSetBstMidEndSepPunct{\mcitedefaultmidpunct}
{\mcitedefaultendpunct}{\mcitedefaultseppunct}\relax
\EndOfBibitem
\bibitem[He \latin{et~al.}(2022)He, Rabe, and Wolverton]{he2022computationally}
He,~J.; Rabe,~K.~M.; Wolverton,~C. Computationally accelerated discovery of functional and structural Heusler materials. \emph{MRS Bulletin} \textbf{2022}, \emph{47}, 559--572\relax
\mciteBstWouldAddEndPuncttrue
\mciteSetBstMidEndSepPunct{\mcitedefaultmidpunct}
{\mcitedefaultendpunct}{\mcitedefaultseppunct}\relax
\EndOfBibitem
\bibitem[Jia \latin{et~al.}(2022)Jia, Deng, Bao, Yao, Li, Li, Chen, Wang, Mao, Cao, Sui, Wu, Wang, Zhang, and Liu]{jia2022unsupervised}
Jia,~X.; Deng,~Y.; Bao,~X.; Yao,~H.; Li,~S.; Li,~Z.; Chen,~C.; Wang,~X.; Mao,~J.; Cao,~F.; Sui,~J.; Wu,~J.; Wang,~C.; Zhang,~Q.; Liu,~X. Unsupervised machine learning for discovery of promising half-Heusler thermoelectric materials. \emph{npj Computational Materials} \textbf{2022}, \emph{8}, 34\relax
\mciteBstWouldAddEndPuncttrue
\mciteSetBstMidEndSepPunct{\mcitedefaultmidpunct}
{\mcitedefaultendpunct}{\mcitedefaultseppunct}\relax
\EndOfBibitem
\bibitem[Kranthiraja and Saeki(2022)Kranthiraja, and Saeki]{kranthiraja2022machine}
Kranthiraja,~K.; Saeki,~A. Machine Learning-Assisted Polymer Design for Improving the Performance of Non-Fullerene Organic Solar Cells. \emph{ACS Applied Materials \& Interfaces} \textbf{2022}, \emph{14}, 28936--28944\relax
\mciteBstWouldAddEndPuncttrue
\mciteSetBstMidEndSepPunct{\mcitedefaultmidpunct}
{\mcitedefaultendpunct}{\mcitedefaultseppunct}\relax
\EndOfBibitem
\bibitem[Sun \latin{et~al.}(2019)Sun, Zheng, Yang, Zhang, Shah, Wu, Sun, Feng, Chen, Xiao, Lu, Li, and Sun]{sun2019machine}
Sun,~W.; Zheng,~Y.; Yang,~K.; Zhang,~Q.; Shah,~A.~A.; Wu,~Z.; Sun,~Y.; Feng,~L.; Chen,~D.; Xiao,~Z.; Lu,~S.; Li,~Y.; Sun,~K. Machine learning--assisted molecular design and efficiency prediction for high-performance organic photovoltaic materials. \emph{Science advances} \textbf{2019}, \emph{5}, eaay4275\relax
\mciteBstWouldAddEndPuncttrue
\mciteSetBstMidEndSepPunct{\mcitedefaultmidpunct}
{\mcitedefaultendpunct}{\mcitedefaultseppunct}\relax
\EndOfBibitem
\bibitem[Yang \latin{et~al.}(2022)Yang, Tao, He, McCutcheon, and Li]{yang2022machine}
Yang,~J.; Tao,~L.; He,~J.; McCutcheon,~J.~R.; Li,~Y. Machine learning enables interpretable discovery of innovative polymers for gas separation membranes. \emph{Science Advances} \textbf{2022}, \emph{8}, eabn9545\relax
\mciteBstWouldAddEndPuncttrue
\mciteSetBstMidEndSepPunct{\mcitedefaultmidpunct}
{\mcitedefaultendpunct}{\mcitedefaultseppunct}\relax
\EndOfBibitem
\bibitem[MacLeod \latin{et~al.}(2020)MacLeod, Parlane, Morrissey, H{\"a}se, Roch, Dettelbach, Moreira, Yunker, Rooney, Deeth, \latin{et~al.} others]{macleod2020self}
MacLeod,~B.~P.; Parlane,~F.~G.; Morrissey,~T.~D.; H{\"a}se,~F.; Roch,~L.~M.; Dettelbach,~K.~E.; Moreira,~R.; Yunker,~L.~P.; Rooney,~M.~B.; Deeth,~J.~R.; others Self-driving laboratory for accelerated discovery of thin-film materials. \emph{Science Advances} \textbf{2020}, \emph{6}, eaaz8867\relax
\mciteBstWouldAddEndPuncttrue
\mciteSetBstMidEndSepPunct{\mcitedefaultmidpunct}
{\mcitedefaultendpunct}{\mcitedefaultseppunct}\relax
\EndOfBibitem
\bibitem[Khatamsaz \latin{et~al.}(2022)Khatamsaz, Vela, Singh, Johnson, Allaire, and Arr{\'o}yave]{khatamsaz2022multi}
Khatamsaz,~D.; Vela,~B.; Singh,~P.; Johnson,~D.~D.; Allaire,~D.; Arr{\'o}yave,~R. Multi-objective materials bayesian optimization with active learning of design constraints: Design of ductile refractory multi-principal-element alloys. \emph{Acta Materialia} \textbf{2022}, \emph{236}, 118133\relax
\mciteBstWouldAddEndPuncttrue
\mciteSetBstMidEndSepPunct{\mcitedefaultmidpunct}
{\mcitedefaultendpunct}{\mcitedefaultseppunct}\relax
\EndOfBibitem
\bibitem[Kim \latin{et~al.}(2019)Kim, Chandrasekaran, Jha, and Ramprasad]{kim2019active}
Kim,~C.; Chandrasekaran,~A.; Jha,~A.; Ramprasad,~R. Active-learning and materials design: the example of high glass transition temperature polymers. \emph{Mrs Communications} \textbf{2019}, \emph{9}, 860--866\relax
\mciteBstWouldAddEndPuncttrue
\mciteSetBstMidEndSepPunct{\mcitedefaultmidpunct}
{\mcitedefaultendpunct}{\mcitedefaultseppunct}\relax
\EndOfBibitem
\bibitem[Rohr \latin{et~al.}(2020)Rohr, Stein, Guevarra, Wang, Haber, Aykol, Suram, and Gregoire]{rohr2020benchmarking}
Rohr,~B.; Stein,~H.~S.; Guevarra,~D.; Wang,~Y.; Haber,~J.~A.; Aykol,~M.; Suram,~S.~K.; Gregoire,~J.~M. Benchmarking the acceleration of materials discovery by sequential learning. \emph{Chemical science} \textbf{2020}, \emph{11}, 2696--2706\relax
\mciteBstWouldAddEndPuncttrue
\mciteSetBstMidEndSepPunct{\mcitedefaultmidpunct}
{\mcitedefaultendpunct}{\mcitedefaultseppunct}\relax
\EndOfBibitem
\bibitem[Wang \latin{et~al.}(2022)Wang, Liang, McDannald, Takeuchi, and Kusne]{wang2022benchmarking}
Wang,~A.; Liang,~H.; McDannald,~A.; Takeuchi,~I.; Kusne,~A.~G. Benchmarking active learning strategies for materials optimization and discovery. \emph{Oxford Open Materials Science} \textbf{2022}, \emph{2}, itac006\relax
\mciteBstWouldAddEndPuncttrue
\mciteSetBstMidEndSepPunct{\mcitedefaultmidpunct}
{\mcitedefaultendpunct}{\mcitedefaultseppunct}\relax
\EndOfBibitem
\bibitem[Borg \latin{et~al.}(2023)Borg, Muckley, Nyby, Saal, Ward, Mehta, and Meredig]{borg2023quantifying}
Borg,~C.~K.; Muckley,~E.~S.; Nyby,~C.; Saal,~J.~E.; Ward,~L.; Mehta,~A.; Meredig,~B. Quantifying the performance of machine learning models in materials discovery. \emph{Digital Discovery} \textbf{2023}, \emph{2}, 327--338\relax
\mciteBstWouldAddEndPuncttrue
\mciteSetBstMidEndSepPunct{\mcitedefaultmidpunct}
{\mcitedefaultendpunct}{\mcitedefaultseppunct}\relax
\EndOfBibitem
\bibitem[Duros \latin{et~al.}(2017)Duros, Grizou, Xuan, Hosni, Long, Miras, and Cronin]{humanrobot2017}
Duros,~V.; Grizou,~J.; Xuan,~W.; Hosni,~Z.; Long,~D.-L.; Miras,~H.~N.; Cronin,~L. Human versus robots in the discovery and crystallization of gigantic polyoxometalates. \emph{Angewandte Chemie} \textbf{2017}, \emph{129}, 10955--10960\relax
\mciteBstWouldAddEndPuncttrue
\mciteSetBstMidEndSepPunct{\mcitedefaultmidpunct}
{\mcitedefaultendpunct}{\mcitedefaultseppunct}\relax
\EndOfBibitem
\bibitem[Duros \latin{et~al.}(2019)Duros, Grizou, Sharma, Mehr, Bubliauskas, Frei, Miras, and Cronin]{humanrobot2019}
Duros,~V.; Grizou,~J.; Sharma,~A.; Mehr,~S. H.~M.; Bubliauskas,~A.; Frei,~P.; Miras,~H.~N.; Cronin,~L. Intuition-Enabled Machine Learning Beats the Competition When Joint Human-Robot Teams Perform Inorganic Chemical Experiments. \emph{Journal of Chemical Information and Modeling} \textbf{2019}, \emph{59}, 2664--2671, PMID: 31025861\relax
\mciteBstWouldAddEndPuncttrue
\mciteSetBstMidEndSepPunct{\mcitedefaultmidpunct}
{\mcitedefaultendpunct}{\mcitedefaultseppunct}\relax
\EndOfBibitem
\bibitem[Shields \latin{et~al.}(2021)Shields, Stevens, Li, Parasram, Damani, Alvarado, Janey, Adams, and Doyle]{shields2021bayesian}
Shields,~B.~J.; Stevens,~J.; Li,~J.; Parasram,~M.; Damani,~F.; Alvarado,~J. I.~M.; Janey,~J.~M.; Adams,~R.~P.; Doyle,~A.~G. Bayesian reaction optimization as a tool for chemical synthesis. \emph{Nature} \textbf{2021}, \emph{590}, 89--96\relax
\mciteBstWouldAddEndPuncttrue
\mciteSetBstMidEndSepPunct{\mcitedefaultmidpunct}
{\mcitedefaultendpunct}{\mcitedefaultseppunct}\relax
\EndOfBibitem
\bibitem[Fu \latin{et~al.}(2019)Fu, Wang, and Sun]{fu2019polymer}
Fu,~H.; Wang,~Z.; Sun,~Y. Polymer donors for high-performance non-fullerene organic solar cells. \emph{Angew. Chem. Int} \textbf{2019}, \emph{58}, 4442--4453\relax
\mciteBstWouldAddEndPuncttrue
\mciteSetBstMidEndSepPunct{\mcitedefaultmidpunct}
{\mcitedefaultendpunct}{\mcitedefaultseppunct}\relax
\EndOfBibitem
\bibitem[Nagasawa \latin{et~al.}(2018)Nagasawa, Al-Naamani, and Saeki]{nagasawa2018computer}
Nagasawa,~S.; Al-Naamani,~E.; Saeki,~A. Computer-aided screening of conjugated polymers for organic solar cell: classification by random forest. \emph{J. Phys. Chem} \textbf{2018}, \emph{9}, 2639--2646\relax
\mciteBstWouldAddEndPuncttrue
\mciteSetBstMidEndSepPunct{\mcitedefaultmidpunct}
{\mcitedefaultendpunct}{\mcitedefaultseppunct}\relax
\EndOfBibitem
\bibitem[Miyake and Saeki(2021)Miyake, and Saeki]{miyake2021machine}
Miyake,~Y.; Saeki,~A. Machine learning-assisted development of organic solar cell materials: issues, analyses, and outlooks. \emph{The Journal of Physical Chemistry Letters} \textbf{2021}, \emph{12}, 12391--12401\relax
\mciteBstWouldAddEndPuncttrue
\mciteSetBstMidEndSepPunct{\mcitedefaultmidpunct}
{\mcitedefaultendpunct}{\mcitedefaultseppunct}\relax
\EndOfBibitem
\bibitem[Greenstein and Hutchison(2023)Greenstein, and Hutchison]{greenstein2023screening}
Greenstein,~B.~L.; Hutchison,~G.~R. Screening Efficient Tandem Organic Solar Cells with Machine Learning and Genetic Algorithms. \emph{The Journal of Physical Chemistry C} \textbf{2023}, \emph{127}, 6179--6191\relax
\mciteBstWouldAddEndPuncttrue
\mciteSetBstMidEndSepPunct{\mcitedefaultmidpunct}
{\mcitedefaultendpunct}{\mcitedefaultseppunct}\relax
\EndOfBibitem
\bibitem[Lopez \latin{et~al.}(2016)Lopez, Pyzer-Knapp, Simm, Lutzow, Li, Seress, Hachmann, and Aspuru-Guzik]{lopez2016harvard}
Lopez,~S.~A.; Pyzer-Knapp,~E.~O.; Simm,~G.~N.; Lutzow,~T.; Li,~K.; Seress,~L.~R.; Hachmann,~J.; Aspuru-Guzik,~A. The Harvard organic photovoltaic dataset. \emph{Scientific data} \textbf{2016}, \emph{3}, 1--7\relax
\mciteBstWouldAddEndPuncttrue
\mciteSetBstMidEndSepPunct{\mcitedefaultmidpunct}
{\mcitedefaultendpunct}{\mcitedefaultseppunct}\relax
\EndOfBibitem
\bibitem[Tshitoyan \latin{et~al.}(2019)Tshitoyan, Dagdelen, Weston, Dunn, Rong, Kononova, Persson, Ceder, and Jain]{tshitoyan2019unsupervised}
Tshitoyan,~V.; Dagdelen,~J.; Weston,~L.; Dunn,~A.; Rong,~Z.; Kononova,~O.; Persson,~K.~A.; Ceder,~G.; Jain,~A. Unsupervised word embeddings capture latent knowledge from materials science literature. \emph{Nature} \textbf{2019}, \emph{571}, 95--98\relax
\mciteBstWouldAddEndPuncttrue
\mciteSetBstMidEndSepPunct{\mcitedefaultmidpunct}
{\mcitedefaultendpunct}{\mcitedefaultseppunct}\relax
\EndOfBibitem
\bibitem[Kononova \latin{et~al.}(2019)Kononova, Huo, He, Rong, Botari, Sun, Tshitoyan, and Ceder]{kononova2019text}
Kononova,~O.; Huo,~H.; He,~T.; Rong,~Z.; Botari,~T.; Sun,~W.; Tshitoyan,~V.; Ceder,~G. Text-mined dataset of inorganic materials synthesis recipes. \emph{Scientific data} \textbf{2019}, \emph{6}, 1--11\relax
\mciteBstWouldAddEndPuncttrue
\mciteSetBstMidEndSepPunct{\mcitedefaultmidpunct}
{\mcitedefaultendpunct}{\mcitedefaultseppunct}\relax
\EndOfBibitem
\bibitem[Guo \latin{et~al.}(2021)Guo, Ibanez-Lopez, Gao, Quach, Coley, Jensen, and Barzilay]{guo2021automated}
Guo,~J.; Ibanez-Lopez,~A.~S.; Gao,~H.; Quach,~V.; Coley,~C.~W.; Jensen,~K.~F.; Barzilay,~R. Automated Chemical Reaction Extraction from Scientific Literature. \emph{Journal of Chemical Information and Modeling} \textbf{2021}, \relax
\mciteBstWouldAddEndPunctfalse
\mciteSetBstMidEndSepPunct{\mcitedefaultmidpunct}
{}{\mcitedefaultseppunct}\relax
\EndOfBibitem
\bibitem[Court and Cole(2018)Court, and Cole]{court2018auto}
Court,~C.~J.; Cole,~J.~M. Auto-generated materials database of Curie and N{\'e}el temperatures via semi-supervised relationship extraction. \emph{Sci. Data} \textbf{2018}, \emph{5}, 1--12\relax
\mciteBstWouldAddEndPuncttrue
\mciteSetBstMidEndSepPunct{\mcitedefaultmidpunct}
{\mcitedefaultendpunct}{\mcitedefaultseppunct}\relax
\EndOfBibitem
\bibitem[Jensen \latin{et~al.}(2019)Jensen, Kim, Kwon, Gani, Roman-Leshkov, Moliner, Corma, and Olivetti]{jensen2019machine}
Jensen,~Z.; Kim,~E.; Kwon,~S.; Gani,~T.~Z.; Roman-Leshkov,~Y.; Moliner,~M.; Corma,~A.; Olivetti,~E. A machine learning approach to zeolite synthesis enabled by automatic literature data extraction. \emph{ACS central science} \textbf{2019}, \emph{5}, 892--899\relax
\mciteBstWouldAddEndPuncttrue
\mciteSetBstMidEndSepPunct{\mcitedefaultmidpunct}
{\mcitedefaultendpunct}{\mcitedefaultseppunct}\relax
\EndOfBibitem
\bibitem[Shetty and Ramprasad(2020)Shetty, and Ramprasad]{shettyautomated}
Shetty,~P.; Ramprasad,~R. Automated knowledge extraction from polymer literature using natural language processing. \emph{iScience} \textbf{2020}, \emph{24}, 101922\relax
\mciteBstWouldAddEndPuncttrue
\mciteSetBstMidEndSepPunct{\mcitedefaultmidpunct}
{\mcitedefaultendpunct}{\mcitedefaultseppunct}\relax
\EndOfBibitem
\bibitem[Shetty and Ramprasad(2021)Shetty, and Ramprasad]{shetty2021machine}
Shetty,~P.; Ramprasad,~R. Machine-Guided Polymer Knowledge Extraction Using Natural Language Processing: The Example of Named Entity Normalization. \emph{Journal of Chemical Information and Modeling} \textbf{2021}, \emph{61}, 5377--5385\relax
\mciteBstWouldAddEndPuncttrue
\mciteSetBstMidEndSepPunct{\mcitedefaultmidpunct}
{\mcitedefaultendpunct}{\mcitedefaultseppunct}\relax
\EndOfBibitem
\bibitem[Shetty \latin{et~al.}(2023)Shetty, Rajan, Kuenneth, Gupta, Panchumarti, Holm, Zhang, and Ramprasad]{shetty2023general}
Shetty,~P.; Rajan,~A.~C.; Kuenneth,~C.; Gupta,~S.; Panchumarti,~L.~P.; Holm,~L.; Zhang,~C.; Ramprasad,~R. A general-purpose material property data extraction pipeline from large polymer corpora using natural language processing. \emph{npj Computational Materials} \textbf{2023}, \emph{9}, 52\relax
\mciteBstWouldAddEndPuncttrue
\mciteSetBstMidEndSepPunct{\mcitedefaultmidpunct}
{\mcitedefaultendpunct}{\mcitedefaultseppunct}\relax
\EndOfBibitem
\bibitem[Weininger \latin{et~al.}(1989)Weininger, Weininger, and Weininger]{weininger1989smiles}
Weininger,~D.; Weininger,~A.; Weininger,~J.~L. SMILES. 2. Algorithm for generation of unique SMILES notation. \emph{J Chem Inf Comput Sci} \textbf{1989}, \emph{29}, 97--101\relax
\mciteBstWouldAddEndPuncttrue
\mciteSetBstMidEndSepPunct{\mcitedefaultmidpunct}
{\mcitedefaultendpunct}{\mcitedefaultseppunct}\relax
\EndOfBibitem
\bibitem[Lee \latin{et~al.}(2020)Lee, Jeong, Kim, Lee, Kang, Woo, and Kim]{lee2020eco}
Lee,~S.; Jeong,~D.; Kim,~C.; Lee,~C.; Kang,~H.; Woo,~H.~Y.; Kim,~B.~J. Eco-friendly polymer solar cells: Advances in green-solvent processing and material design. \emph{Acs Nano} \textbf{2020}, \emph{14}, 14493--14527\relax
\mciteBstWouldAddEndPuncttrue
\mciteSetBstMidEndSepPunct{\mcitedefaultmidpunct}
{\mcitedefaultendpunct}{\mcitedefaultseppunct}\relax
\EndOfBibitem
\bibitem[Sorrentino \latin{et~al.}(2021)Sorrentino, Kozma, Luzzati, and Po]{sorrentino2021interlayers}
Sorrentino,~R.; Kozma,~E.; Luzzati,~S.; Po,~R. Interlayers for non-fullerene based polymer solar cells: distinctive features and challenges. \emph{Energy \& Environmental Science} \textbf{2021}, \emph{14}, 180--223\relax
\mciteBstWouldAddEndPuncttrue
\mciteSetBstMidEndSepPunct{\mcitedefaultmidpunct}
{\mcitedefaultendpunct}{\mcitedefaultseppunct}\relax
\EndOfBibitem
\bibitem[Jin \latin{et~al.}(2023)Jin, Wang, Ma, Shen, Belfiore, Bao, and Tang]{jin2023recent}
Jin,~J.; Wang,~Q.; Ma,~K.; Shen,~W.; Belfiore,~L.~A.; Bao,~X.; Tang,~J. Recent developments of polymer solar cells with photovoltaic performance over 17\%. \emph{Advanced Functional Materials} \textbf{2023}, \emph{33}, 2213324\relax
\mciteBstWouldAddEndPuncttrue
\mciteSetBstMidEndSepPunct{\mcitedefaultmidpunct}
{\mcitedefaultendpunct}{\mcitedefaultseppunct}\relax
\EndOfBibitem
\bibitem[Yin \latin{et~al.}(2020)Yin, Yan, Hu, Ho, Zhan, Li, and So]{yin2020recent}
Yin,~H.; Yan,~C.; Hu,~H.; Ho,~J. K.~W.; Zhan,~X.; Li,~G.; So,~S.~K. Recent progress of all-polymer solar cells--From chemical structure and device physics to photovoltaic performance. \emph{Materials Science and Engineering: R: Reports} \textbf{2020}, \emph{140}, 100542\relax
\mciteBstWouldAddEndPuncttrue
\mciteSetBstMidEndSepPunct{\mcitedefaultmidpunct}
{\mcitedefaultendpunct}{\mcitedefaultseppunct}\relax
\EndOfBibitem
\bibitem[Ma \latin{et~al.}(2022)Ma, Zhang, Feng, and Guo]{ma2022polymer}
Ma,~S.; Zhang,~H.; Feng,~K.; Guo,~X. Polymer Acceptors for High-Performance All-Polymer Solar Cells. \emph{Chemistry--A European Journal} \textbf{2022}, \emph{28}, e202200222\relax
\mciteBstWouldAddEndPuncttrue
\mciteSetBstMidEndSepPunct{\mcitedefaultmidpunct}
{\mcitedefaultendpunct}{\mcitedefaultseppunct}\relax
\EndOfBibitem
\bibitem[Kuenneth and Ramprasad(2023)Kuenneth, and Ramprasad]{kuenneth2023polybert}
Kuenneth,~C.; Ramprasad,~R. polyBERT: a chemical language model to enable fully machine-driven ultrafast polymer informatics. \emph{Nature Communications} \textbf{2023}, \emph{14}, 4099\relax
\mciteBstWouldAddEndPuncttrue
\mciteSetBstMidEndSepPunct{\mcitedefaultmidpunct}
{\mcitedefaultendpunct}{\mcitedefaultseppunct}\relax
\EndOfBibitem
\bibitem[Zhao \latin{et~al.}(2020)Zhao, Yao, Ali, Miao, and Meng]{zhao2020recent}
Zhao,~J.; Yao,~C.; Ali,~M.~U.; Miao,~J.; Meng,~H. Recent advances in high-performance organic solar cells enabled by acceptor--donor--acceptor--donor--acceptor (A--DA' D--A) type acceptors. \emph{Materials Chemistry Frontiers} \textbf{2020}, \emph{4}, 3487--3504\relax
\mciteBstWouldAddEndPuncttrue
\mciteSetBstMidEndSepPunct{\mcitedefaultmidpunct}
{\mcitedefaultendpunct}{\mcitedefaultseppunct}\relax
\EndOfBibitem
\bibitem[Padula and Troisi(2019)Padula, and Troisi]{padula2019concurrent}
Padula,~D.; Troisi,~A. Concurrent optimization of organic donor--acceptor pairs through machine learning. \emph{Advanced Energy Materials} \textbf{2019}, \emph{9}, 1902463\relax
\mciteBstWouldAddEndPuncttrue
\mciteSetBstMidEndSepPunct{\mcitedefaultmidpunct}
{\mcitedefaultendpunct}{\mcitedefaultseppunct}\relax
\EndOfBibitem
\bibitem[Doan~Tran \latin{et~al.}(2020)Doan~Tran, Kim, Chen, Chandrasekaran, Batra, Venkatram, Kamal, Lightstone, Gurnani, Shetty, Ramprasad, Laws, Shelton, and Ramprasad]{pg}
Doan~Tran,~H.; Kim,~C.; Chen,~L.; Chandrasekaran,~A.; Batra,~R.; Venkatram,~S.; Kamal,~D.; Lightstone,~J.~P.; Gurnani,~R.; Shetty,~P.; Ramprasad,~M.; Laws,~J.; Shelton,~M.; Ramprasad,~R. Machine-learning predictions of polymer properties with Polymer Genome. \emph{J. Appl. Phys.} \textbf{2020}, \emph{128}, 171104\relax
\mciteBstWouldAddEndPuncttrue
\mciteSetBstMidEndSepPunct{\mcitedefaultmidpunct}
{\mcitedefaultendpunct}{\mcitedefaultseppunct}\relax
\EndOfBibitem
\bibitem[Gurnani \latin{et~al.}(2023)Gurnani, Kuenneth, Toland, and Ramprasad]{gurnani2023polymer}
Gurnani,~R.; Kuenneth,~C.; Toland,~A.; Ramprasad,~R. Polymer Informatics at Scale with Multitask Graph Neural Networks. \emph{Chemistry of Materials} \textbf{2023}, \emph{35}, 1560--1567\relax
\mciteBstWouldAddEndPuncttrue
\mciteSetBstMidEndSepPunct{\mcitedefaultmidpunct}
{\mcitedefaultendpunct}{\mcitedefaultseppunct}\relax
\EndOfBibitem
\bibitem[Kuenneth \latin{et~al.}(2021)Kuenneth, Schertzer, and Ramprasad]{kuenneth2021copolymer}
Kuenneth,~C.; Schertzer,~W.; Ramprasad,~R. Copolymer informatics with multitask deep neural networks. \emph{Macromolecules} \textbf{2021}, \emph{54}, 5957--5961\relax
\mciteBstWouldAddEndPuncttrue
\mciteSetBstMidEndSepPunct{\mcitedefaultmidpunct}
{\mcitedefaultendpunct}{\mcitedefaultseppunct}\relax
\EndOfBibitem
\bibitem[Kamath \latin{et~al.}(2018)Kamath, Vargas-Hern{\'a}ndez, Krems, Carrington, and Manzhos]{kamath2018neural}
Kamath,~A.; Vargas-Hern{\'a}ndez,~R.~A.; Krems,~R.~V.; Carrington,~T.; Manzhos,~S. Neural networks vs Gaussian process regression for representing potential energy surfaces: A comparative study of fit quality and vibrational spectrum accuracy. \emph{The Journal of chemical physics} \textbf{2018}, \emph{148}\relax
\mciteBstWouldAddEndPuncttrue
\mciteSetBstMidEndSepPunct{\mcitedefaultmidpunct}
{\mcitedefaultendpunct}{\mcitedefaultseppunct}\relax
\EndOfBibitem
\bibitem[Pedregosa \latin{et~al.}(2011)Pedregosa, Varoquaux, Gramfort, Michel, Thirion, Grisel, Blondel, Prettenhofer, Weiss, Dubourg, Vanderplas, Passos, Cournapeau, Brucher, Perrot, and Duchesnay]{scikit-learn}
Pedregosa,~F. \latin{et~al.}  Scikit-learn: Machine Learning in {P}ython. \emph{Journal of Machine Learning Research} \textbf{2011}, \emph{12}, 2825--2830\relax
\mciteBstWouldAddEndPuncttrue
\mciteSetBstMidEndSepPunct{\mcitedefaultmidpunct}
{\mcitedefaultendpunct}{\mcitedefaultseppunct}\relax
\EndOfBibitem
\bibitem[Williams and Rasmussen(2006)Williams, and Rasmussen]{williams2006gaussian}
Williams,~C.~K.; Rasmussen,~C.~E. \emph{Gaussian processes for machine learning}; MIT press Cambridge, MA, 2006; Vol.~2\relax
\mciteBstWouldAddEndPuncttrue
\mciteSetBstMidEndSepPunct{\mcitedefaultmidpunct}
{\mcitedefaultendpunct}{\mcitedefaultseppunct}\relax
\EndOfBibitem
\bibitem[Noack \latin{et~al.}(2020)Noack, Doerk, Li, Streit, Vaia, Yager, and Fukuto]{noack2020autonomous}
Noack,~M.~M.; Doerk,~G.~S.; Li,~R.; Streit,~J.~K.; Vaia,~R.~A.; Yager,~K.~G.; Fukuto,~M. Autonomous materials discovery driven by Gaussian process regression with inhomogeneous measurement noise and anisotropic kernels. \emph{Scientific reports} \textbf{2020}, \emph{10}, 17663\relax
\mciteBstWouldAddEndPuncttrue
\mciteSetBstMidEndSepPunct{\mcitedefaultmidpunct}
{\mcitedefaultendpunct}{\mcitedefaultseppunct}\relax
\EndOfBibitem
\bibitem[Lookman \latin{et~al.}(2019)Lookman, Balachandran, Xue, and Yuan]{lookman2019active}
Lookman,~T.; Balachandran,~P.~V.; Xue,~D.; Yuan,~R. Active learning in materials science with emphasis on adaptive sampling using uncertainties for targeted design. \emph{npj Computational Materials} \textbf{2019}, \emph{5}, 21\relax
\mciteBstWouldAddEndPuncttrue
\mciteSetBstMidEndSepPunct{\mcitedefaultmidpunct}
{\mcitedefaultendpunct}{\mcitedefaultseppunct}\relax
\EndOfBibitem
\bibitem[Tran \latin{et~al.}(2020)Tran, Neiswanger, Yoon, Zhang, Xing, and Ulissi]{tran2020methods}
Tran,~K.; Neiswanger,~W.; Yoon,~J.; Zhang,~Q.; Xing,~E.; Ulissi,~Z.~W. Methods for comparing uncertainty quantifications for material property predictions. \emph{Machine Learning: Science and Technology} \textbf{2020}, \emph{1}, 025006\relax
\mciteBstWouldAddEndPuncttrue
\mciteSetBstMidEndSepPunct{\mcitedefaultmidpunct}
{\mcitedefaultendpunct}{\mcitedefaultseppunct}\relax
\EndOfBibitem
\bibitem[Srinivas \latin{et~al.}(2009)Srinivas, Krause, Kakade, and Seeger]{srinivas2009gaussian}
Srinivas,~N.; Krause,~A.; Kakade,~S.~M.; Seeger,~M. Gaussian process optimization in the bandit setting: No regret and experimental design. \emph{arXiv preprint arXiv:0912.3995} \textbf{2009}, \relax
\mciteBstWouldAddEndPunctfalse
\mciteSetBstMidEndSepPunct{\mcitedefaultmidpunct}
{}{\mcitedefaultseppunct}\relax
\EndOfBibitem
\bibitem[Jones \latin{et~al.}(1998)Jones, Schonlau, and Welch]{jones1998efficient}
Jones,~D.~R.; Schonlau,~M.; Welch,~W.~J. Efficient global optimization of expensive black-box functions. \emph{Journal of Global optimization} \textbf{1998}, \emph{13}, 455--492\relax
\mciteBstWouldAddEndPuncttrue
\mciteSetBstMidEndSepPunct{\mcitedefaultmidpunct}
{\mcitedefaultendpunct}{\mcitedefaultseppunct}\relax
\EndOfBibitem
\bibitem[Agrawal and Goyal(2013)Agrawal, and Goyal]{agrawal2013thompson}
Agrawal,~S.; Goyal,~N. Thompson sampling for contextual bandits with linear payoffs. International conference on machine learning. 2013; pp 127--135\relax
\mciteBstWouldAddEndPuncttrue
\mciteSetBstMidEndSepPunct{\mcitedefaultmidpunct}
{\mcitedefaultendpunct}{\mcitedefaultseppunct}\relax
\EndOfBibitem
\bibitem[Bin \latin{et~al.}(2022)Bin, van~der Pol, Li, van Gorkom, Wienk, and Janssen]{bin2022efficient}
Bin,~H.; van~der Pol,~T.~P.; Li,~J.; van Gorkom,~B.~T.; Wienk,~M.~M.; Janssen,~R.~A. Efficient organic solar cells with small energy losses based on a wide-bandgap trialkylsilyl-substituted donor polymer and a non-fullerene acceptor. \emph{Chemical Engineering Journal} \textbf{2022}, \emph{435}, 134878\relax
\mciteBstWouldAddEndPuncttrue
\mciteSetBstMidEndSepPunct{\mcitedefaultmidpunct}
{\mcitedefaultendpunct}{\mcitedefaultseppunct}\relax
\EndOfBibitem
\bibitem[Jain \latin{et~al.}(2020)Jain, van Zuylen, Hajishirzi, and Beltagy]{jain2020scirex}
Jain,~S.; van Zuylen,~M.; Hajishirzi,~H.; Beltagy,~I. SciREX: A challenge dataset for document-level information extraction. \emph{arXiv preprint arXiv:2005.00512} \textbf{2020}, \relax
\mciteBstWouldAddEndPunctfalse
\mciteSetBstMidEndSepPunct{\mcitedefaultmidpunct}
{}{\mcitedefaultseppunct}\relax
\EndOfBibitem
\bibitem[Song \latin{et~al.}(2020)Song, Ham, Noh, Lee, and Kang]{song2020efficiency}
Song,~C.~E.; Ham,~H.; Noh,~J.; Lee,~S.~K.; Kang,~I.-N. Efficiency enhancement of a fluorinated wide-bandgap polymer for ternary nonfullerene organic solar cells. \emph{Polymer} \textbf{2020}, \emph{188}, 122131\relax
\mciteBstWouldAddEndPuncttrue
\mciteSetBstMidEndSepPunct{\mcitedefaultmidpunct}
{\mcitedefaultendpunct}{\mcitedefaultseppunct}\relax
\EndOfBibitem
\bibitem[Wang \latin{et~al.}(2020)Wang, Liu, Zhou, Mo, Han, Lai, Chen, Zheng, Zhu, Xie, and He]{wang2020bromination}
Wang,~H.; Liu,~T.; Zhou,~J.; Mo,~D.; Han,~L.; Lai,~H.; Chen,~H.; Zheng,~N.; Zhu,~Y.; Xie,~Z.; He,~F. Bromination: an alternative strategy for non-fullerene small molecule acceptors. \emph{Advanced Science} \textbf{2020}, \emph{7}, 1903784\relax
\mciteBstWouldAddEndPuncttrue
\mciteSetBstMidEndSepPunct{\mcitedefaultmidpunct}
{\mcitedefaultendpunct}{\mcitedefaultseppunct}\relax
\EndOfBibitem
\bibitem[He \latin{et~al.}(2019)He, Shahid, Wu, Jiao, Eisner, Hodsden, Fei, Anthopoulos, McNeill, Durrant, and Heeney]{he2019fused}
He,~Q.; Shahid,~M.; Wu,~J.; Jiao,~X.; Eisner,~F.~D.; Hodsden,~T.; Fei,~Z.; Anthopoulos,~T.~D.; McNeill,~C.~R.; Durrant,~J.~R.; Heeney,~M. Fused cyclopentadithienothiophene acceptor enables ultrahigh short-circuit current and high efficiency \textgreater 11\% in as-cast organic solar cells. \emph{Advanced Functional Materials} \textbf{2019}, \emph{29}, 1904956\relax
\mciteBstWouldAddEndPuncttrue
\mciteSetBstMidEndSepPunct{\mcitedefaultmidpunct}
{\mcitedefaultendpunct}{\mcitedefaultseppunct}\relax
\EndOfBibitem
\bibitem[Li \latin{et~al.}(2020)Li, Cheng, Zhang, Yang, and Liu]{li2020higher}
Li,~G.; Cheng,~H.; Zhang,~Y.; Yang,~T.; Liu,~Y. Higher open circuit voltage caused by chlorinated polymers endows improved efficiency of binary organic solar cell. \emph{Organic Electronics} \textbf{2020}, \emph{83}, 105776\relax
\mciteBstWouldAddEndPuncttrue
\mciteSetBstMidEndSepPunct{\mcitedefaultmidpunct}
{\mcitedefaultendpunct}{\mcitedefaultseppunct}\relax
\EndOfBibitem
\bibitem[Huang \latin{et~al.}(2019)Huang, Peng, Xie, Song, Hong, Chen, Wei, and Ge]{huang2019novel}
Huang,~J.; Peng,~R.; Xie,~L.; Song,~W.; Hong,~L.; Chen,~S.; Wei,~Q.; Ge,~Z. A novel polymer donor based on dithieno [2, 3-d: 2', 3'-d''] benzo [1, 2-b: 4, 5-b'] dithiophene for highly efficient polymer solar cells. \emph{Journal of Materials Chemistry A} \textbf{2019}, \emph{7}, 2646--2652\relax
\mciteBstWouldAddEndPuncttrue
\mciteSetBstMidEndSepPunct{\mcitedefaultmidpunct}
{\mcitedefaultendpunct}{\mcitedefaultseppunct}\relax
\EndOfBibitem
\bibitem[Firdaus \latin{et~al.}(2017)Firdaus, Maffei, Cruciani, M{\"u}ller, Liu, Lopatin, Wehbe, Ndjawa, Amassian, Laquai, and Beaujuge]{firdaus2017polymer}
Firdaus,~Y.; Maffei,~L.~P.; Cruciani,~F.; M{\"u}ller,~M.~A.; Liu,~S.; Lopatin,~S.; Wehbe,~N.; Ndjawa,~G. O.~N.; Amassian,~A.; Laquai,~F.; Beaujuge,~P.~M. Polymer Main-Chain Substitution Effects on the Efficiency of Nonfullerene BHJ Solar Cells. \emph{Advanced Energy Materials} \textbf{2017}, \emph{7}, 1700834\relax
\mciteBstWouldAddEndPuncttrue
\mciteSetBstMidEndSepPunct{\mcitedefaultmidpunct}
{\mcitedefaultendpunct}{\mcitedefaultseppunct}\relax
\EndOfBibitem
\bibitem[Zhu \latin{et~al.}(2019)Zhu, Zhong, Qiu, Lyu, Zhou, Zhang, Song, Xu, Wang, Ali, Feng, Shi, Gu, Ying, Zhang, and Liu]{zhu2019aggregation}
Zhu,~L. \latin{et~al.}  Aggregation-induced multilength scaled morphology enabling 11.76\% efficiency in all-polymer solar cells using printing fabrication. \emph{Advanced Materials} \textbf{2019}, \emph{31}, 1902899\relax
\mciteBstWouldAddEndPuncttrue
\mciteSetBstMidEndSepPunct{\mcitedefaultmidpunct}
{\mcitedefaultendpunct}{\mcitedefaultseppunct}\relax
\EndOfBibitem
\bibitem[Li \latin{et~al.}(2020)Li, Weng, Ryu, Guo, Zhang, Xia, Fu, Wei, Min, Zhang, Woo, and Sun]{li2020non}
Li,~X.; Weng,~K.; Ryu,~H.~S.; Guo,~J.; Zhang,~X.; Xia,~T.; Fu,~H.; Wei,~D.; Min,~J.; Zhang,~Y.; Woo,~H.~Y.; Sun,~Y. Non-Fullerene Organic Solar Cells Based on Benzo [1, 2-b: 4, 5-b'] difuran-Conjugated Polymer with 14\% Efficiency. \emph{Advanced Functional Materials} \textbf{2020}, \emph{30}, 1906809\relax
\mciteBstWouldAddEndPuncttrue
\mciteSetBstMidEndSepPunct{\mcitedefaultmidpunct}
{\mcitedefaultendpunct}{\mcitedefaultseppunct}\relax
\EndOfBibitem
\bibitem[Ma \latin{et~al.}(2020)Ma, Zhang, Yao, Xu, Wang, Zu, and Hou]{ma2020high}
Ma,~L.; Zhang,~S.; Yao,~H.; Xu,~Y.; Wang,~J.; Zu,~Y.; Hou,~J. High-efficiency nonfullerene organic solar cells enabled by 1000 nm thick active layers with a low trap-state density. \emph{ACS applied materials \& interfaces} \textbf{2020}, \emph{12}, 18777--18784\relax
\mciteBstWouldAddEndPuncttrue
\mciteSetBstMidEndSepPunct{\mcitedefaultmidpunct}
{\mcitedefaultendpunct}{\mcitedefaultseppunct}\relax
\EndOfBibitem
\bibitem[Sun \latin{et~al.}(2019)Sun, Pan, Chen, Wang, Sun, Shang, Qiu, Min, Lv, Meng, \latin{et~al.} others]{sun2019achieving}
Sun,~C.; Pan,~F.; Chen,~S.; Wang,~R.; Sun,~R.; Shang,~Z.; Qiu,~B.; Min,~J.; Lv,~M.; Meng,~L.; others Achieving fast charge separation and low nonradiative recombination loss by rational fluorination for high-efficiency polymer solar cells. \emph{Advanced Materials} \textbf{2019}, \emph{31}, 1905480\relax
\mciteBstWouldAddEndPuncttrue
\mciteSetBstMidEndSepPunct{\mcitedefaultmidpunct}
{\mcitedefaultendpunct}{\mcitedefaultseppunct}\relax
\EndOfBibitem
\bibitem[Zhang \latin{et~al.}(2020)Zhang, Song, Liu, Zhang, Qu, Yang, Yuan, Mahmood, Liu, He, Baran, and Wang]{zhang2020electron}
Zhang,~C.; Song,~X.; Liu,~K.-K.; Zhang,~M.; Qu,~J.; Yang,~C.; Yuan,~G.-Z.; Mahmood,~A.; Liu,~F.; He,~F.; Baran,~D.; Wang,~J.-L. Electron-Deficient and Quinoid Central Unit Engineering for Unfused Ring-Based A1--D--A2--D--A1-Type Acceptor Enables High Performance Nonfullerene Polymer Solar Cells with High Voc and PCE Simultaneously. \emph{Small} \textbf{2020}, \emph{16}, 1907681\relax
\mciteBstWouldAddEndPuncttrue
\mciteSetBstMidEndSepPunct{\mcitedefaultmidpunct}
{\mcitedefaultendpunct}{\mcitedefaultseppunct}\relax
\EndOfBibitem
\bibitem[He \latin{et~al.}(2020)He, Zheng, Lu, Guo, Gao, Pang, Mola, Zhao, and Zhang]{he2020highly}
He,~E.; Zheng,~Z.; Lu,~Y.; Guo,~F.; Gao,~S.; Pang,~X.; Mola,~G.~T.; Zhao,~L.; Zhang,~Y. Highly efficient non-fullerene polymer solar cells from a benzo [1, 2-b: 4, 5-b'] difuran-based conjugated polymer with improved stabilities. \emph{Journal of Materials Chemistry A} \textbf{2020}, \emph{8}, 11381--11390\relax
\mciteBstWouldAddEndPuncttrue
\mciteSetBstMidEndSepPunct{\mcitedefaultmidpunct}
{\mcitedefaultendpunct}{\mcitedefaultseppunct}\relax
\EndOfBibitem
\bibitem[Fujinuma \latin{et~al.}(2022)Fujinuma, DeCost, Hattrick-Simpers, and Lofland]{fujinuma2022big}
Fujinuma,~N.; DeCost,~B.; Hattrick-Simpers,~J.; Lofland,~S.~E. Why big data and compute are not necessarily the path to big materials science. \emph{Communications Materials} \textbf{2022}, \emph{3}, 59\relax
\mciteBstWouldAddEndPuncttrue
\mciteSetBstMidEndSepPunct{\mcitedefaultmidpunct}
{\mcitedefaultendpunct}{\mcitedefaultseppunct}\relax
\EndOfBibitem
\bibitem[Dyakonov(2004)]{dyakonov2004electrical}
Dyakonov,~V. Electrical aspects of operation of polymer--fullerene solar cells. \emph{Thin Solid Films} \textbf{2004}, \emph{451}, 493--497\relax
\mciteBstWouldAddEndPuncttrue
\mciteSetBstMidEndSepPunct{\mcitedefaultmidpunct}
{\mcitedefaultendpunct}{\mcitedefaultseppunct}\relax
\EndOfBibitem
\bibitem[McInnes \latin{et~al.}(2018)McInnes, Healy, and Melville]{mcinnes2018umap}
McInnes,~L.; Healy,~J.; Melville,~J. Umap: Uniform manifold approximation and projection for dimension reduction. \emph{arXiv preprint arXiv:1802.03426} \textbf{2018}, \relax
\mciteBstWouldAddEndPunctfalse
\mciteSetBstMidEndSepPunct{\mcitedefaultmidpunct}
{}{\mcitedefaultseppunct}\relax
\EndOfBibitem
\bibitem[Qian \latin{et~al.}(2023)Qian, Guo, Tu, Li, Coley, and Barzilay]{molscribe}
Qian,~Y.; Guo,~J.; Tu,~Z.; Li,~Z.; Coley,~C.~W.; Barzilay,~R. MolScribe: Robust Molecular Structure Recognition with Image-to-Graph Generation. \emph{Journal of Chemical Information and Modeling} \textbf{2023}, \emph{63}, 1925--1934\relax
\mciteBstWouldAddEndPuncttrue
\mciteSetBstMidEndSepPunct{\mcitedefaultmidpunct}
{\mcitedefaultendpunct}{\mcitedefaultseppunct}\relax
\EndOfBibitem
\bibitem[Filippov and Nicklaus(2009)Filippov, and Nicklaus]{osra}
Filippov,~I.~V.; Nicklaus,~M.~C. Optical Structure Recognition Software To Recover Chemical Information: OSRA, An Open Source Solution. \emph{Journal of Chemical Information and Modeling} \textbf{2009}, \emph{49}, 740--743, PMID: 19434905\relax
\mciteBstWouldAddEndPuncttrue
\mciteSetBstMidEndSepPunct{\mcitedefaultmidpunct}
{\mcitedefaultendpunct}{\mcitedefaultseppunct}\relax
\EndOfBibitem
\bibitem[Yang and Ding(2007)Yang, and Ding]{yang2007novel}
Yang,~Q.; Ding,~S. Novel algorithm to calculate hypervolume indicator of Pareto approximation set. International Conference on Intelligent Computing. 2007; pp 235--244\relax
\mciteBstWouldAddEndPuncttrue
\mciteSetBstMidEndSepPunct{\mcitedefaultmidpunct}
{\mcitedefaultendpunct}{\mcitedefaultseppunct}\relax
\EndOfBibitem
\bibitem[Emmerich \latin{et~al.}(2011)Emmerich, Deutz, and Klinkenberg]{emmerich2011hypervolume}
Emmerich,~M.~T.; Deutz,~A.~H.; Klinkenberg,~J.~W. Hypervolume-based expected improvement: Monotonicity properties and exact computation. 2011 IEEE Congress of Evolutionary Computation (CEC). 2011; pp 2147--2154\relax
\mciteBstWouldAddEndPuncttrue
\mciteSetBstMidEndSepPunct{\mcitedefaultmidpunct}
{\mcitedefaultendpunct}{\mcitedefaultseppunct}\relax
\EndOfBibitem
\bibitem[Xu \latin{et~al.}(2018)Xu, Fukuda, Karki, Park, Kimura, Jinno, Watanabe, Yamamoto, Shimomura, Kitazawa, Yokota, Umezu, Nguyen, and Someya]{xu2018thermally}
Xu,~X.; Fukuda,~K.; Karki,~A.; Park,~S.; Kimura,~H.; Jinno,~H.; Watanabe,~N.; Yamamoto,~S.; Shimomura,~S.; Kitazawa,~D.; Yokota,~T.; Umezu,~S.; Nguyen,~T.-Q.; Someya,~T. Thermally stable, highly efficient, ultraflexible organic photovoltaics. \emph{Proceedings of the National Academy of Sciences} \textbf{2018}, \emph{115}, 4589--4594\relax
\mciteBstWouldAddEndPuncttrue
\mciteSetBstMidEndSepPunct{\mcitedefaultmidpunct}
{\mcitedefaultendpunct}{\mcitedefaultseppunct}\relax
\EndOfBibitem
\bibitem[Gevorgyan \latin{et~al.}(2017)Gevorgyan, Heckler, Bundgaard, Corazza, H{\"o}sel, S{\o}ndergaard, dos Reis~Benatto, J{\o}rgensen, and Krebs]{gevorgyan2017improving}
Gevorgyan,~S.~A.; Heckler,~I.~M.; Bundgaard,~E.; Corazza,~M.; H{\"o}sel,~M.; S{\o}ndergaard,~R.~R.; dos Reis~Benatto,~G.~A.; J{\o}rgensen,~M.; Krebs,~F.~C. Improving, characterizing and predicting the lifetime of organic photovoltaics. \emph{Journal of Physics D: Applied Physics} \textbf{2017}, \emph{50}, 103001\relax
\mciteBstWouldAddEndPuncttrue
\mciteSetBstMidEndSepPunct{\mcitedefaultmidpunct}
{\mcitedefaultendpunct}{\mcitedefaultseppunct}\relax
\EndOfBibitem
\bibitem[Rafique \latin{et~al.}(2018)Rafique, Abdullah, Sulaiman, and Iwamoto]{rafique2018fundamentals}
Rafique,~S.; Abdullah,~S.~M.; Sulaiman,~K.; Iwamoto,~M. Fundamentals of bulk heterojunction organic solar cells: An overview of stability/degradation issues and strategies for improvement. \emph{Renewable and Sustainable Energy Reviews} \textbf{2018}, \emph{84}, 43--53\relax
\mciteBstWouldAddEndPuncttrue
\mciteSetBstMidEndSepPunct{\mcitedefaultmidpunct}
{\mcitedefaultendpunct}{\mcitedefaultseppunct}\relax
\EndOfBibitem
\end{mcitethebibliography}

\end{document}